\documentclass[preprint,3p,12pt]{elsarticle}
\usepackage{mathrsfs}
\usepackage{amsmath}
\usepackage{stmaryrd}
\usepackage{bbding}
\usepackage{dcolumn}
\usepackage{graphicx}
\usepackage{amsfonts}
\usepackage{amssymb}
\usepackage{psfrag}
\usepackage{wrapfig}
\usepackage{subfigure}
\usepackage{makeidx}
\usepackage{bm}
\usepackage{epsf}
\usepackage{color}
\usepackage{epsfig}
\usepackage{setspace}
\usepackage{graphicx}
\usepackage{epstopdf}
\usepackage{psfrag}
\usepackage{subfigure}
\usepackage{multirow}
\usepackage{diagbox}
\usepackage{verbatim}

\usepackage{makecell}
\usepackage{float}
\usepackage{esint}
\usepackage{comment}
\usepackage{movie15}
\usepackage{hyperref}
\usepackage[ruled,linesnumbered]{algorithm2e}

\newcommand{\mathsym}[1]{{}}
\newcommand{\unicode}[1]{{}}
\begin{document}
	
\title{A p-multigrid compact gas-kinetic scheme for steady-state acceleration}
	
	\author[HKUST1]{Xing Ji}
	\ead{xjiad@connect.ust.hk}
		
	\author[HKUST2]{Wei Shyy}
	\ead{weishyy@ust.hk}
	
	\author[HKUST1,HKUST2,HKUST3]{Kun Xu\corref{cor1}}
	\ead{makxu@ust.hk}

	\address[HKUST1]{Department of Mathematics, Hong Kong University of Science and Technology, Clear Water Bay, Kowloon, Hong Kong}
	\address[HKUST2]{Department of Mechanical and Aerospace Engineering, Hong Kong University of Science and Technology, Clear Water Bay, Kowloon, Hong Kong}
	\address[HKUST3]{Shenzhen Research Institute, Hong Kong University of Science and Technology, Shenzhen, China}
	\cortext[cor1]{Corresponding author}

\begin{abstract}
	
In this paper, the high-order compact gas-kinetic scheme (CGKS) on three-dimensional hybrid unstructured mesh is further developed with the p-multigrid technique for steady-state solution acceleration.
The p-multigrid strategy is a two-level algorithm. On the high-order level, the high-order CGKS is used to evolve both cell-averaged conservative flow variables and their gradients under high-order compact initial reconstruction at the beginning of next time step.
On the low-order level, starting from the high-order level solution the cell-averaged conservative flow variables is evolved by a first-order scheme, where implicit backward Euler smoother is adopted for accelerating the convergence of steady-state solution.
The final iterative updating scheme becomes numerically simple and computationally efficient.
The effectiveness of the p-multigrid method is validated in both subsonic and supersonic flow simulations in two- and three-dimensional space
 with hybrid unstructured mesh.
One order of magnitude speedup in the convergence rate has be achieved by the approach in comparison with the explicit counterpart.
\end{abstract}


\maketitle

\section{Introduction}
In the past decades, the development of high-order methods for computational fluid dynamics has made great progress \cite{wang2013high}.
The representative algorithms include the weighted essentially non-oscillatory (WENO) method \cite{antoniadis2017assessment}, the Discontinuous Galerkin (DG)
method \cite{hindenlang2012dg},
the reconstructed-DG (rDG) method \cite{liu2017reconstructed}, the boundary variation diminishing (BVD) method \cite{sun2016BVD}, and the variational finite  volume (VFV) method \cite{wang2017compact}.

In recent years, a class of high-order compact gas-kinetic scheme (CGKS) \cite{pan2016unstructuredcompact,ji2020hweno} has been developed from the second-order gas-kinetic scheme (GKS) \cite{GKS-2001}.
The time-dependent interface evolution solution in CGKS is based on an analytical integral solution of the Bhatnagar--Gross--Krook equation \cite{BGK}, which recovers a transition process from the kinetic scale flux vector splitting to the central difference Lax-Wendroff type discretization.
The time-accurate gas distribution function can be used to evaluate both flux function across a cell interface and the time accurate flow variables at the cell interface.
As a result, besides updating the cell-averaged flow variables as a conventional finite volume scheme,
the cell-averaged gradients of flow variables can be updated simultaneously through the Gauss-Green theorem.
Based on the flow variables and their gradients, a compact high-order reconstruction scheme can be properly designed.
Another distinguishable feature of GKS is the use of the explicit two-stage fourth-order temporal discretization (S2O4)
or the multi-stage multi-derivative time marching scheme for the high-order temporal discretization \cite{li2016twostage,multi-derivative}.
In comparison with the fourth-order four-stage Runge-Kutta (RK) method, the S2O4 is more efficient with only two stages \cite{ji2021compact}.
The limiting process for the possible discontinuous flux function in time has been studied as well in the CGKS \cite{zhao2021direct}.
Due to the physically reliable evolution process and efficient time discretization, the CGKS achieves remarkable success for unsteady compressible flow simulation, such as in computational aeroacoustics \cite{zhao2019acoustic} and implicit large eddy simulation \cite{ji2021compact}.
The CGKS has been constructed up to the eighth-order of accuracy in space on structured mesh \cite{zhao2019compact}, and fourth-order accuracy on triangular mesh with the possible use of a large CFL number, such as $CFL \approx 0.8$ \cite{zhao2020compact,zhao2021direct}.
The computation of hypersonic flow around a space vehicle on 3-D hybrid mesh demonstrates its robustness in flow simulation with complex geometry \cite{ji2021gradient}.
However, in large-scale simulation, the mesh size around the boundary and far field can be different in thousands of times.
The rate of convergence for the steady state can be extremely slow using the explicit time marching approach.
Thus, it becomes necessary to develop the acceleration technique for CGKS.

The implicit temporal discretization and multigrid strategy are two popular and efficient techniques for speeding up convergence.
Implicit high-order methods have been developed based on the lower-upper symmetric Gauss-Seidel (LU-SGS) method \cite{antoniadis2017assessment} and Generalized minimal residual (GMRES) method \cite{yang2019robust}.
The implicit methods for GKS and unified GKS (UGKS) have also been constructed \cite{tan2017time,li2017dualtime,zhu2016implicit}.
These methods solve a linear system of equations derived from the linearization of a fully implicit scheme in each iteration.
However, due to the complicated nonlinear evolution model in updating cell interface values, for the high-order compact GKS it is not trivial to derive the linearized Jacobian matrix for the cell-averaged gradients in the development of implicit CGKS.
On the other hand, a series of p-multigrid methods have been developed for compact methods with multiple degrees of freedom in each cell \cite{ronquist1987spectral,luo2006pmultigrid,bassi2009high,liang2009pmultigrid}.
The discretized system is solved by
iterating the approximate solutions on different order accuracy recursively.
For instance, to solve a DG scheme with polynomial order of 2 inside each element, a three-level algorithm with polynomial orders of $p$=2, 1, 0 can be adopted and the solution is iterated on each level accuracy.
The implicit and low-storage iterative method can be used on the low-order level, e.g., $p$=0, to eliminate the low-frequency error efficiently.
The explicit time marching method is used on the high-order level, e.g., $p$=2, and avoids the expensive memory consumption of the Jacobian matrix evaluation for high-order solution \cite{luo2006pmultigrid}.

In this paper, an efficient and low-storage p-multigrid method is developed for accelerating the CGKS in steady-state simulation.
Following the idea in \cite{luo2006pmultigrid}, a two-level V-cycle algorithm is designed.
On the high-order level, the original explicit CGKS is adopted and both the cell-averaged conservative flow variables their gradients, i.e., $\overline{\textbf{W}}$ and  $\overline{\textbf{W}}_{x_i}$, $i=1,2,3$, are updated.
Then the flow solution and residual is transferred to the low-order solution, i.e., a piece-wise constant approximation inside each control volume.
On the low-order level, only the cell-averaged conservative flow variables are smoothed, evolved, and transferred back to the high-order level.
The first-order flux function, such as the Riemann solution, is applied in the low-order level and the implicit matrix-free point-relaxation method with local time stepping is used in the time integration \cite{yuan2002comparison}.
The similarity and differences between the p-multigrid method for the CGKS and the DG scheme \cite{luo2006pmultigrid} are as follows. \\
i) With $\overline{\textbf{W}}$ and $\overline{\textbf{W}}_{x_i}$, the CGKS has a similar memory cost as the $P1$-DG case on high-order approximation level. \\
ii) Since $\overline{\textbf{W}}_{x_i}$ takes no effect on the transformation of
$\overline{\textbf{W}}$ between the two levels, the restriction and prolongation operator for CGKS are very simple, i.e., directly copy the values of $\overline{\textbf{W}}$ and the corresponding residuals. \\
iii) The explicit multi-stage RK method is adopted for the DG scheme while the single-step second-order time-marching method can be used for the CGKS. \\
The computational cost between the explicit and p-multigrid iteration is compared through the computations for nearly incompressible and supersonic flows.
Moreover, the influence of different fluxes on the low-order level is investigated.
The kinetic vector flux splitting solver and classical approximate Riemann solver are compared.
 Due to the implicit method on the low-order level, the parallel performance for the scheme will be evaluated.

This paper is organized as follows.
The basic ingredients of the high-order CGKS on mixed-element mesh are introduced in Section 2.
In Section 3, a p-multigrid approach for CGKS is presented.
Numerical examples including both inviscid and viscous flow computations are given in Section 4.
The last section is the conclusion.

\section{Compact gas-kinetic scheme framework}

The 3-D gas-kinetic BGK equation \cite{BGK} is
\begin{equation}\label{bgk}
f_t+\textbf{u}\cdot\nabla f=\frac{g-f}{\tau},
\end{equation}
where $f=f(\textbf{x},t,\textbf{u},\xi)$ is the gas distribution function, which is a function of space $\textbf{x}$, time $t$, particle velocity $\textbf{u}$, and internal variable $\xi$.
$g$ is the equilibrium state approached by $f$
and $\tau$ is the collision time.

The collision term satisfies the compatibility condition
\begin{equation*}\label{compatibility}
\int \frac{g-f}{\tau} \pmb{\psi} \text{d}\Xi=0,
\end{equation*}
where $\pmb{\psi}=(1,\textbf{u},\displaystyle \frac{1}{2}(\textbf{u}^2+\xi^2))^T$,
$\text{d}\Xi=\text{d}u_1\text{d}u_2\text{d}u_3\text{d}\xi_1...\text{d}\xi_{K}$,
$K$ is the number of internal degrees of freedom, i.e.
$K=(5-3\gamma)/(\gamma-1)$ in 3-D case, and $\gamma$
is the specific heat ratio.

In the continuum flow regime with the smoothness assumption, based on the Chapman-Enskog expansion the gas distribution function can be expressed as \cite{xu2014direct} ,
\begin{align*}
f=g-\tau D_{\textbf{u}}g+\tau D_{\textbf{u}}(\tau
D_{\textbf{u}})g-\tau D_{\textbf{u}}[\tau D_{\textbf{u}}(\tau
D_{\textbf{u}})g]+...,
\end{align*}
where $D_{\textbf{u}}={\partial}/{\partial t}+\textbf{u}\cdot \nabla$.
Different hydrodynamic equations can be derived by truncating on different orders of $\tau$.
With the first-order truncation, i.e.,
\begin{align*} \label{ce-ns}
f=g-\tau (\textbf{u} \cdot \nabla g + g_t),
\end{align*}
the N-S equations can be obtained,
\begin{equation*}\label{ns-conservation}
\begin{split}
\textbf{W}_t+ \nabla \cdot \textbf{F}(\textbf{W},\nabla \textbf{W} )=0,
\end{split}
\end{equation*}
with $\tau = \mu / p$ and $Pr=1$.

The conservative flow variables and their fluxes are the moments of the gas distribution function
\begin{align}\label{point}
\textbf{W}(\textbf{x},t)=\int \pmb{\psi} f(\textbf{x},t,\textbf{u},\xi)\text{d}\Xi,
\end{align}
and
\begin{equation}\label{f-to-flux}
\textbf{F}(\textbf{x},t)=
\int \textbf{u} \pmb{\psi} f(\textbf{x},t,\textbf{u},\xi)\text{d}\Xi.
\end{equation}

\subsection{Finite volume discretization on hybrid grids}

For a 3-D polyhedral cell $\Omega_i$, the boundary can be expressed as
\begin{equation*}
\partial \Omega_i=\bigcup_{p=1}^{N_f}\Gamma_{ip},
\end{equation*}
where $N_f$ is the number of cell interfaces for cell $\Omega_i$.
$N_f=4$ for tetrahedron, $N_f=5$ for prism and pyramid, $N_f=6$ for hexahedron.
The semi-discretized form of finite volume method for conservation laws can be written as
\begin{equation}\label{semidiscrete}
\frac{\text{d} \textbf{W}_{i}}{\text{d}t}=\mathcal{L}(\textbf{W}_i)=-\frac{1}{\left| \Omega_i \right|} \sum_{p=1}^{N_f} \int_{\Gamma_{ip}}
\textbf{F}(\textbf{W}(\textbf{x},t))\cdot\textbf{n}_p \text{d}s,
\end{equation}
with
\begin{equation*}\label{f-to-flux-in-normal-direction}
\textbf{F}(\textbf{W}(\textbf{x},t))\cdot \textbf{n}_p=\int\pmb{\psi}  f(\textbf{x},t,\textbf{u},\xi) \textbf{u}\cdot \textbf{n}_p \text{d}\Xi,
\end{equation*}
where $\textbf{W}_{i}$ is the cell averaged values over cell $\Omega_i$, $\left|
\Omega_i \right|$ is the volume of $\Omega_i$, $\textbf{F}$ is the interface fluxes, and $\textbf{n}_p=(n_1,n_2,n_3)^T$ is the unit vector representing the outer normal direction of $\Gamma_{ip}$.
Through the iso-parametric transformation,
the Gaussian quadrature points can be determined and $\textbf{F}_{ip}(t)$ can be approximated by the numerical quadrature
\begin{equation*}\label{fv-3d-general-quadrature}
\sum_{p=1}^{N_f} \int_{\Gamma_{ip}}
\textbf{F}(\textbf{W}(\textbf{x},t))\cdot\textbf{n}_p \text{d}s =  \left|\Gamma_{ip}\right| \sum_{k=1}^{M} \omega_k
\textbf{F}(\textbf{x}_{p,k},t)\cdot\textbf{n}_p.
\end{equation*}
In this work, the linear element is considered.
To meet the requirement of a third-order spatial accuracy,
three Gaussian points are used for a triangular face and four Gaussian points are used for a quadrilateral face.
In the computation, the fluxes are obtained under the local coordinates.
The details can be found in \cite{pan2020high,ji2021two}.

\subsection{Gas-kinetic solver}
Based on the integral solution of BGK equation, a second-order time accurate gas distribution function at a local Gaussian point $\textbf{x}=(0,0,0)$ is constructed as
\begin{align}\label{2nd-flux}
f(\textbf{0},t,\textbf{u},\xi)
=&(1-e^{-t/\tau_n}) g^{c}+[(t+\tau)e^{-t/\tau_n}-\tau]a_{x_i}^{c}u_i g^{c}\nonumber
+(t-\tau+\tau e^{-t/\tau_n})A^{c}  g^{c}\nonumber\\
+&e^{-t/\tau_n}g^l[1-(\tau+t)a_{x_i}^{l}u_i-\tau A^l]H(u_1)\nonumber\\
+&e^{-t/\tau_n}g^r[1-(\tau+t)a_{x_i}^{r}u_i-\tau A^r] (1-H(u_1)).
\end{align}
The superscripts $l,r$ represent the initial gas distribution function $f_0$ at the left and right sides of a cell interface.
The superscript $c$ is the corresponding
equilibrium state $g$ in space and time.
The integral solution basically states a physical process from the particle free transport in $f_0$ in the kinetic scale
to the hydrodynamic flow evolution in the integral of $g$ term.
The flow evolution at the cell interface depends on the ratio of time step
to the  local particle collision time $\Delta t/\tau$.

The $g^k,~k=l,r$ has a form of a Maxwell distribution
\begin{align*}
g^k=\rho^k (\frac{\lambda^k}{\pi})e^{-\lambda^k((u_i-U_i)^2+\xi^2)},
\end{align*}
which can be fully determined from the
 macroscopic variables $\textbf{W}
^l, \textbf{W}
^r$ through spatial reconstruction
\begin{align*}
\int\pmb{\psi} g^{l}\text{d}\Xi=\textbf{W}
^l,\int\pmb{\psi} g^{r}\text{d}\Xi=\textbf{W}
^r.
\end{align*}
The spatial and temporal microscopic derivatives are denoted as
\begin{align*}
a_{x_i} \equiv  (\partial g/\partial x_i)/g=g_{x_i}/g,
A \equiv (\partial g/\partial t)/g=g_t/g,
\end{align*}
which is determined by the spatial derivatives of macroscopic flow
variables and the compatibility condition as follows
\begin{align*}
&\langle a_{x_1}\rangle =\frac{\partial \textbf{W} }{\partial x_1}=\textbf{W}_{x_1},
\langle a_{x_2}\rangle =\frac{\partial \textbf{W} }{\partial x_2}=\textbf{W}_{x_2},
\langle a_{x_3}\rangle =\frac{\partial \textbf{W} }{\partial x_3}=\textbf{W}_{x_3},\nonumber\\
&\langle A+a_{x_1}u_1+a_{x_2}u_2+a_{x_3}u_3\rangle=0,
\end{align*}
where $\left\langle ... \right\rangle$ are the moments of a gas distribution function defined by
\begin{align*}
\langle (...) \rangle  = \int \pmb{\psi} (...) g \text{d} \Xi .
\end{align*}

Similarly, the
equilibrium state $ g^{c}$ and its derivatives $a_{x_i}^c, A_{x_i}^c$ are determined by corresponding $\textbf{W}^c, \textbf{W}^c_{x_i}$.
The details for the compact reconstruction of
$\textbf{W}^{l,r,c}, \textbf{W}^{l,r,c}_{x_i}$
can be found in \cite{ji2021gradient}.
The details for calculating each term in the distribution from the corresponding macroscopic flow variable can refer to \cite{xu2014direct}.

For smooth flow, the time dependent solution in Eq.~\eqref{2nd-flux} can be simplified as \cite{GKS-2001},
\begin{align}\label{2nd-smooth-flux}
f(\textbf{0},t,\textbf{u},\xi)= g^{c}-\tau (a^c_{x_{i}} u_i +A^c)g^{c}+A^c g^{c}t,
\end{align}	
under the assumptions of $g^{l,r}=g^c$, $a^{l,r}_{x_i}=a^c_{x_i}$.
The above gas-kinetic solver for smooth flow has less numerical dissipations than the full GKS solver in Eq.~\eqref{2nd-flux}.
In  smooth flow region, the collision time is determined by
\begin{align*}
\tau=\mu/p,
\end{align*}
where $\mu$ is the dynamic viscosity coefficient and $p$ is the pressure at the cell interface.
In order to properly capture the un-resolved discontinuities, additional numerical dissipation is needed.
The physical collision time $\tau$ in the exponential function part can be replaced by a numerical collision time $\tau_n$.
The same $\tau_n$ as that in \cite{ji2021compact} is adopted in this work.

\subsection{Direct evolution of the cell averaged first-order spatial derivatives} \label{slope-section}

As shown in Eq.~\eqref{2nd-flux}, a time evolution solution at a cell interface is provided by the gas-kinetic solver, which is distinguished from the Riemann solvers with a constant solution.
Recall Eq.(\ref{point}), the conservative variables at the Gaussian point  $\textbf{x}_{p,k}$ can be updated through the moments $\pmb{\psi}$
of the gas distribution function,
\begin{equation*}\label{point-interface}
\begin{aligned}
\textbf{W}_{p,k}(t^{n+1})=\int \pmb{\psi} f^n(\textbf{x}_{p,k},t^{n+1},\textbf{u},\xi) \text{d}\Xi,~ k=1,...,M.
\end{aligned}
\end{equation*}

Then, the cell-averaged first-order derivatives within each element at $t^{n+1}$ can be evaluated based on the divergence theorem,

\begin{equation*}\label{gauss-formula}
\begin{aligned}
\nabla \overline{W}^{n+1} \left| \Omega \right|
&=\int_{\Omega} \nabla \overline{W}(t^{n+1}) \text{d}V
=\int_{\partial \Omega} \overline{W}(t^{n+1}) \textbf{n} \text{d}S
= \sum_{p=1}^{N_f}\sum_{k=1}^{M_p} \omega_{p,k} W^{n+1}_{p,k} \textbf{n}_{p,k} \Delta S_p,
\end{aligned}
\end{equation*}
where $\textbf{n}_{p,k}=((n_{1})_{p,k},(n_{2})_{p,k},(n_{3})_{p,k})$ is the outer unit normal direction at each Gaussian point $\textbf{x}_{p,k}$.

\subsection{Explicit one-step temporal discretization}

The one-step second-order (S1O2) temporal discretization is
adopted here for the steady solution.
Following the definition of Eq.\eqref{semidiscrete},
a second-order time-accurate solution for cell-averaged conservative flow variables $\textbf{W}_i$ are updated by
\begin{equation*}\label{s1o2}
\begin{aligned}
\textbf{W}_i^{n+1}=\textbf{W}_i^n+\Delta t\mathcal
{L}(\textbf{W}_i^n)+\frac{1}{2}\Delta t^2\frac{\partial}{\partial
	t}\mathcal{L}(\textbf{W}_i^n),
\end{aligned}
\end{equation*}
where
$\mathcal{L}(\textbf{W}_i^n)$ and $\frac{\partial}{\partial t}\mathcal{L}(\textbf{W}_i^n)$ are given by
\begin{equation*} \label{flux-operator}
\begin{aligned}
\mathcal{L}(\textbf{W}_i^n)&= -\frac{1}{\left| \Omega_i \right|} \sum_{p=1}^{N_f}
\sum_{k=1}^{M} \omega_{p,k} \textbf{F}(\textbf{x}_{p,k},t_n)\cdot \textbf{n}_{p,k},\\
\frac{\partial}{\partial t}\mathcal{L}(\textbf{W}_i^n)&= -\frac{1}{\left| \Omega_i \right|} \sum_{p=1}^{N_f}\sum_{k=1}^{M} \omega_{p,k}
\partial_t \textbf{F}(\textbf{x}_{p,k},t_n)\cdot \textbf{n}_{p,k}.
\end{aligned}
\end{equation*}
The time dependent gas distribution function at a cell interface is updated in a similar way,
\begin{equation*}\label{step-du}
\begin{split}
f^{n+1}=f^n+\Delta tf_t^n.
\end{split}
\end{equation*}
The $f^{n+1}$ are fully determined by Eq.~\eqref{2nd-flux} or Eq.\eqref{2nd-smooth-flux} and the macroscopic flow variables and their fluxes at the cell interface can be obtained simultaneously by Eq.~\eqref{point} and Eq.~\eqref{f-to-flux}. The details can be found in \cite{ji2020hweno}.

\section{P-multigrid for compact gas-kinetic scheme}
A two-level V-cycle p-multigrid algorithm is used to accelerate CGKS for steady-state solutions.
On the first level, i.e., the high-order level, the explicit CGKS is applied.
On the second level, i.e., the low-order level, the first-order finite volume scheme is used to eliminate the long-wavelength errors efficiently.
Specifically, the algorithm adopted in this work is summarized as follows:
\begin{enumerate}
	\item Perform a time-step on the high-order level, obtain the solution $\textbf{W}^{n+1}_{p_1}$
	\item Transfer the solution  $\textbf{W}^{n+1}_{p_1}$ and the correspoding residual to the low-order level: piece-wise constant $p_0$. Since the cell averaged $\bar{\textbf{W}}^{n+1}$ is independent from the cell averaged slopes  $\bar{\textbf{W}}_{x_i}^{n+1}, ~ i=1,2,3$, the solution on the low-order level becomes
	\begin{align*}
	\textbf{W}_{p_0}=\bar{\textbf{W}}^{n+1}.
	\end{align*}
	The residual is the net fluxes for updating $\bar{\textbf{W}}^{n+1}$,
		\begin{align*}
	\textbf{R}(\textbf{W}^{n+1}_{p_1})=\sum \frac{1}{\Delta t}\int_{t^{n+1}}^{t^{n+1}+\Delta t}\textbf{F}(t) \text{d}t.
	\end{align*}
	The residual is transferred to the low-order level
	\begin{align*}
	\textbf{R}_{p_0}=\textbf{R}(\textbf{W}^{n+1}_{p_1}).
	\end{align*}
	\item Compute the force terms on the low-order level,
		\begin{align*}
	\textbf{S}_{p_0}=\textbf{R}_{p_0} - \textbf{R}(\textbf{W}_{p_0}),
	\end{align*}
	where $\textbf{R}(\textbf{W}_{p_0})$ is computed from the first-order scheme.
	Since the gas-kinetic solver is used on the high-order level, it is physically consistent to use a kinetic-type solver on the low-order level.
	In this paper, the first-order kinetic flux vector splitting (KFVS) scheme is used if no specified.
	In addition, the results by the approximate Riemann solver derived from the hydrodynamic equations are also presented to investigate the influence of difference fluxes on the low-order level to the convergence performance.
	\item Solve the steady equation
	 \begin{align}\label{low-order-eq}
	 \textbf{R}(\tilde{\textbf{W}}_{p_0}) + \textbf{S}_{p_0}=0.
	 \end{align}
	 \item  Interpolate the correction $\textbf{C}_{p_0}$ from the low-order level back to the high-order level, and update the solution
	  \begin{align*}
	  \textbf{C}_{p_0} =\tilde{\textbf{W}}_{p_0} -\textbf{W}_{p_0},\\
	  \tilde{\bar{\textbf{W}}}^{n+1}= \bar{\textbf{W}}^{n+1} + \textbf{C}_{p_0}.
	  \end{align*}
\end{enumerate}

In practice, the explicit iteration for Eq.~\eqref{low-order-eq} brings very limited speedup for convergence. Similar conclusions have been drawn in the p-multigrid DG or CPR methods \cite{luo2006pmultigrid,liang2009pmultigrid}.
Instead, the implicit backward Euler method is used as the time discretization, which is introduced as follows.

Discretize the conservation laws with source term by first-order backward Euler in time,
\begin{equation*} \label{implicit-o}
|\Omega_{ i}| \frac{\overline{\mathbf{U}_{ i}}^{n+1}-\overline{\mathbf{U}_{ i}}^{n}}{\Delta t}+\oint_{\partial \Omega} (\mathbf{F}_{n}^{n+1}-\mathbf{F}_{n}^{n}) \text{d} s=-\oint_{\partial \Omega} \mathbf{F}_{n}^{n} \text{d} s +|\Omega_{ i}|\mathbf{S}_n := \mathbf{R}_{i}^{n} \end{equation*}
where $\mathbf{F}_{n} =\mathbf{F}\cdot \textbf{n}$, $\textbf{n}$ is the outer unit normal direction for each interface,
and $\mathbf{F}_{n}^{n}$ is the explicit flux at $t^n$.
Denote $\Delta {\mathbf{U}}=\overline{\mathbf{U}}^{n+1}-\overline{\mathbf{U}}^{n}$,
the equation is simplified as
\begin{equation}\label{implicit}
|\Omega_{ i}| \frac{\Delta {\mathbf{U}_{ i}}}{\Delta t}+\oint_{\partial \Omega} (\mathbf{F}_{n}^{n+1}-\mathbf{F}_{n}^{n}) \text{d} s = \mathbf{R}_{i}^{n} \end{equation}
Linearize the term by first-order Taylor expansion
\begin{equation*}\label{linearize}
\mathbf{F}_{n}^{n+1}-\mathbf{F}_{n}^{n} \approx \left(\frac{\partial \mathbf{F}_{n}}{\partial \mathbf{U}}\right)^{n}\left(\mathbf{U}^{n+1}-\mathbf{U}^{n}\right)=\mathbf{A}^{n} \Delta {\mathbf{U}},
\end{equation*}
where $ \mathbf{A} = \left(\frac{\partial \mathbf{F}_{n}}{\partial \mathbf{U}}\right) = \left(\frac{\partial (\mathbf{F}\cdot \textbf{n}) }{\partial \mathbf{U}}\right) $ is the Jacobian matrix of the normal flux.
The Jacobian matrix of the first-order L-F flux is adopted, which is given by
\begin{equation*} \label{lf-1st-split}
\begin{split}
&(\mathbf{A} \Delta {\mathbf{U}})_{i,p}=\mathbf{A}_{i,p}^{+} \Delta {\mathbf{U}}_{i}+\mathbf{A}_{i,p}^{-} \Delta {\mathbf{U}}_{i,p},\\
&\mathbf{A}^{\pm}= \frac{1}{2}\left(\mathbf{A} \pm \lambda \mathbf{I} \right),
\end{split}
\end{equation*}
where
$\Delta {\mathbf{U}}_{i,p}$ is the increment of the conservative variables in the neighboring cell sharing the same interface $\Gamma_{ip}$.
In the current computation, the modified spectral radius $\lambda$ is adopted
\begin{equation*}
\begin{split}
\lambda = |U_n|+c + 2 \mu \frac{S}{|\Omega|}.
\end{split}
\end{equation*}

Then we have
\begin{equation*}\begin{array}{l}
\oint_{\partial \Omega} \mathbf{A}^{n} \Delta {\mathbf{U}} \textbf{d} s
= \sum_{p=1}^{N_f} \left(\mathbf{A}_{i}^{+} \Delta {\mathbf{U}}_{i}+\mathbf{A}_{i,p}^{-} \Delta {\mathbf{U}}_{i,p}\right)  S_{i,p}.
\end{array}\end{equation*}
Substitute it into Eq.~\eqref{implicit}, we have
\begin{equation}\label{implicit-dis}
\begin{array}{l}
\Delta {\mathbf{U}}_{i}
\left[
\frac{|\Omega_{i}|}{\Delta t}
+\frac{1}{2}\sum_{p=1}^{N_f} {(\lambda S) }_{i,p}
\right]
+\sum_{p=1}^{N_f} \mathbf{A}_{i,p}^{-} \Delta {\mathbf{U}}_{i,p} S_{i,p}
=\mathbf{R}_{i,j}^{n},
\end{array}
\end{equation}
where
\begin{equation*}
{\lambda }_{i+1 / 2,j}= \text{Max} ( {\lambda }_{i},  {\lambda }_{i,p}).
\end{equation*}

The delta flux $\mathbf{A}_{i,p}^{-} \Delta {\mathbf{U}}_{i,p}$ is approximated as
\begin{equation*}
\mathbf{A}_{i,p}^{-} \Delta {\mathbf{U}}_{i,p} \approx \Delta (\mathbf{F}_{n})_{i,p}^{-}=\mathbf{F}_{n}^{-}\left(\mathbf{U}^{n}_{i,p}+\Delta {\mathbf{U}}_{i,p}\right)-\mathbf{F}_{n}^{-}\left(\mathbf{U}^{n}_{i,p}\right).
\end{equation*}

For high Reynolds number flow under highly stretched mesh, the convergence rate can be further accelerated if the local time step for each cell is used.
Eq.~\eqref{implicit-dis}  is rewritten as
\begin{equation}\label{implicit-local}
\begin{array}{l}
\Delta {\mathbf{U}}_{i}
\left[
\frac{|\Omega_{i}|}{\Delta t_i}
+\frac{1}{2}\sum_{p=1}^{N_f} {(\lambda S) }_{i,p}
\right]
+\sum_{p=1}^{N_f} [\mathbf{F}_{n}^{-}\left(\mathbf{U}^{n}_{i,p}+\Delta {\mathbf{U}}_{i,p}\right)-\mathbf{F}_{n}^{-}\left(\mathbf{U}^{n}_{i,p}\right)]
=\mathbf{R}_{i,j}^{n},
\end{array}
\end{equation}
where $\Delta t_{i}$ is the local time step.
To solve the solution matrix formed by Eq.~\eqref{implicit-dis},
efficient matrix-free iteration methods, such as the LU-SGS method \cite{jameson1987lower} and point relaxation method \cite{yuan2002comparison}, can be applied.

In this work, the point relaxation method is applied as shown in algorithm 1.
Six implicit steps are performed on the low-order level.
In each step, the point relaxation method with two sweeps is used.
The OPENMP is applied to achieve the thread-level parallelism.  
For n-thread parallel computation, each sweep for the point-relaxation method is simply divided into n sub-sweeps by the element index. 
If the series sweeps are for all the elements from 1 to $N_{e}$, the parallel sweep for the ith thread is simply from $\frac{i-1}{n} N_{e}$ to $\frac{i}{n} N_{e}$.
Since the communications are non-blocking among each thread, the results will be slightly different for each running.

\begin{algorithm}
	\caption{Point relaxation method}
	\KwIn{Sweep number $N_{s}$}
	\label{point-relaxation}
	Initialization: $\Delta \mathbf{U}_i=0, i=0,...,N_e$, where $N_e$ is the total cell number\;
	\For{j=1,..,$N_{s}$}
	{
		\For{i=1,..,$N_{e}$}
		{
			Update $\Delta \mathbf{U}_i$ by Eq.~\eqref{implicit-local}\;
		}
		\For{i=$N_{e}$,..,1}
		{
			Update $\Delta \mathbf{U}_i$ by Eq.~\eqref{implicit-local}\;
		}
	}
	Update $\mathbf{U}_i^{n+1}=\mathbf{U}_i^{n}+\Delta \mathbf{U}_i$.
\end{algorithm}

\section{Numerical examples} \label{test-case}

In this section, numerical tests will be presented to validate the proposed scheme.
All the simulations are performed on 3-D mesh.
For 2-D test cases, two layers of mesh are generated in z-direction.
The explicit CFL number for the high-order level is taken as 0.3$\sim $0.5 while the local CFL number for the low-order level is taken as 1000.
For subsonic cases in this work, the flow is quite smooth,
so the compact reconstruction with only linear weights and the simplified smooth flux in Eq.~\eqref{2nd-smooth-flux} are adopted for the CGKS.
For the transonic and supersonic cases,
the reconstruction with non-linear  WENO-type weights and the full flux in Eq.~\eqref{2nd-flux} are used.
If no specified, the reconstruction will be performed on the conservative variables,
the KFVS solver and OPENMP parallelism with 8 thread will be adopted on the low-order level.
The non-dimensional density residual is mainly presented to measure the error level, which is given by
\begin{align*}
Res =\frac{\sum_{1}^{N_c} |\rho^{n+1}_i-\rho^{n}_i|}{\sum_{1}^{N_c} \rho^{n}_i},
\end{align*}
where $N_c$ is the total cell number.

\subsection{Subsonic circular cylinder}

\begin{figure}[htp]	
	\centering
	\includegraphics[width=0.4\textwidth]
	{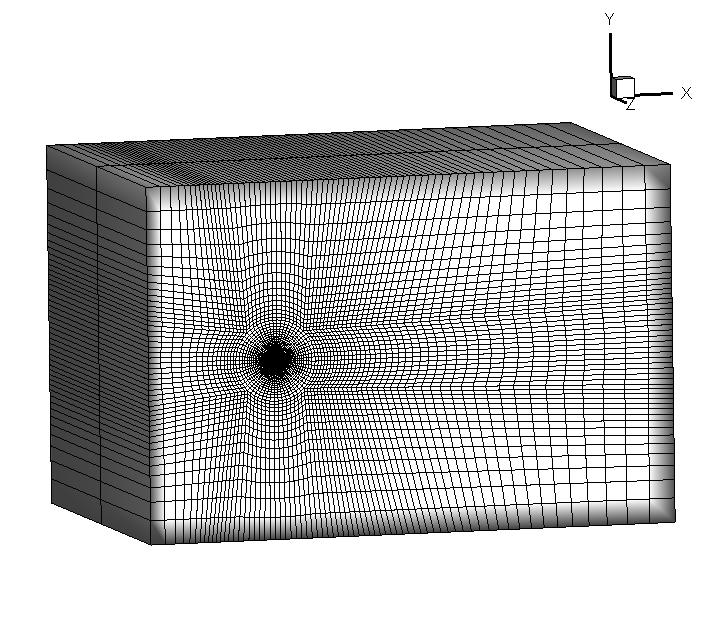}
	\includegraphics[width=0.4\textwidth]
	{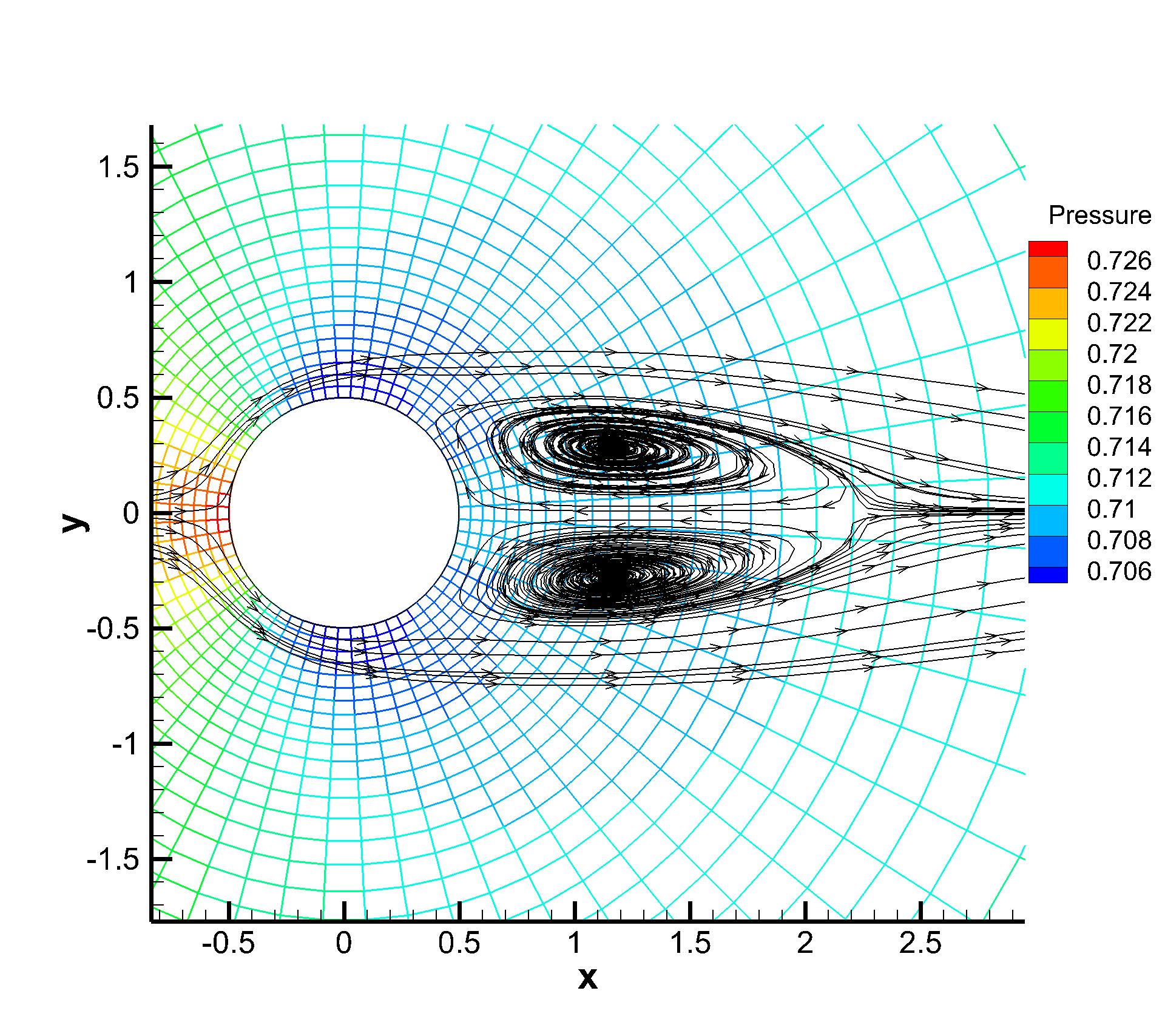}
	\vspace{-4mm} \caption{\label{cylinder-re40-coarse-mesh}
		Circular cylinder: Re=40. Left: Mesh I with $4725 \times 2$ hexahedral cells.
		Right: Local mesh distribution around cylinder colored by pressure and streamline.}
\end{figure}

\begin{figure}[htp]	
	\centering
	\includegraphics[width=0.4\textwidth]
	{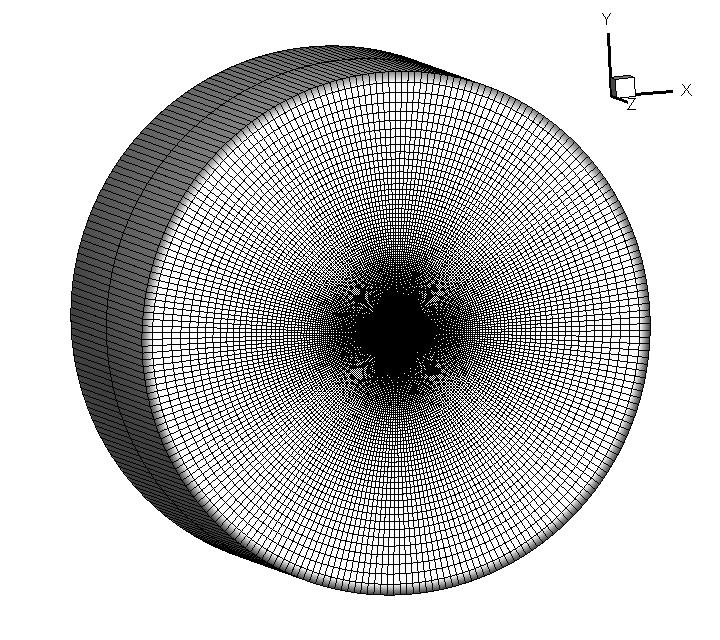}
	\includegraphics[width=0.4\textwidth]
	{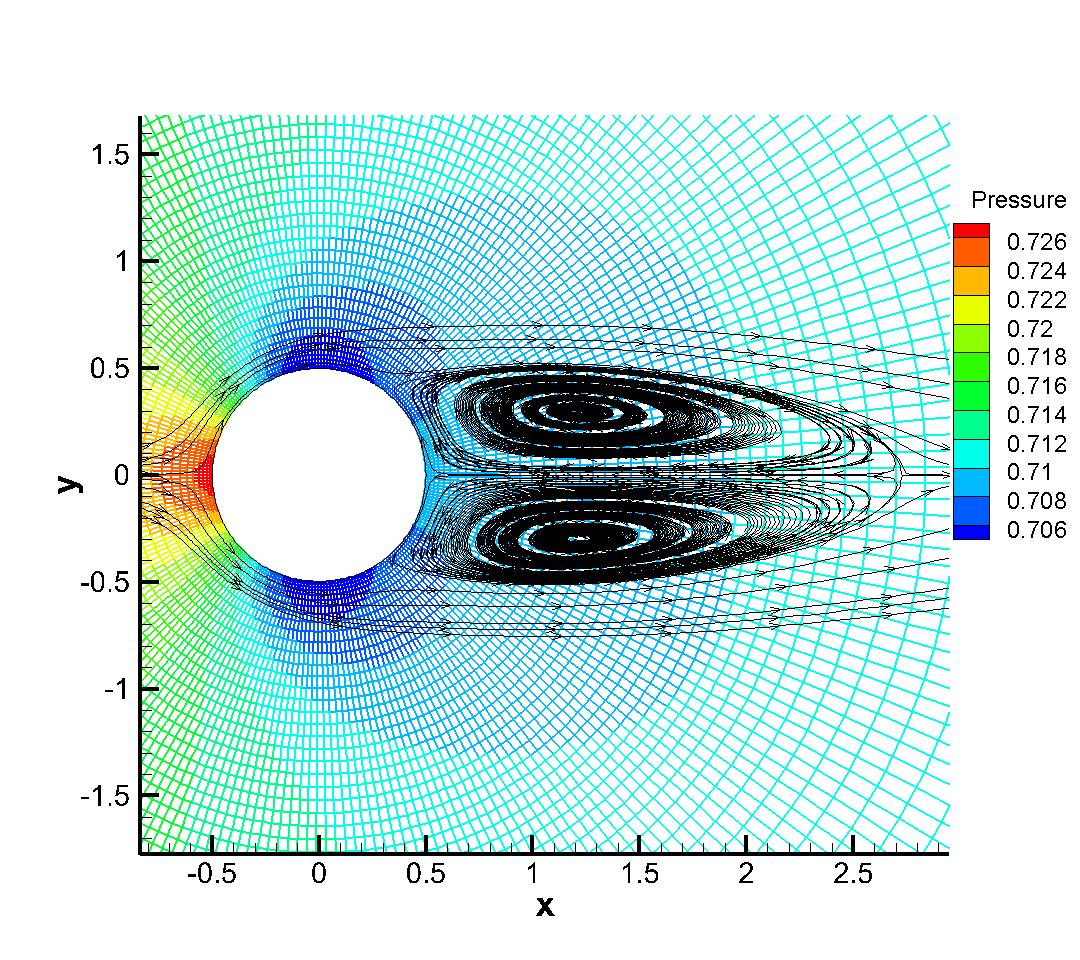}
	\vspace{-4mm} \caption{\label{cylinder-re40-fine-mesh}
		Circular cylinder: Re=40. Left:  Mesh II with $241 \times 114 \times 2$  hexahedral cells.
		Right: Local mesh distribution around cylinder colored by pressure and streamline. }
\end{figure}

\begin{figure}[htp]	
	\centering
	\includegraphics[width=0.32\textwidth]
	{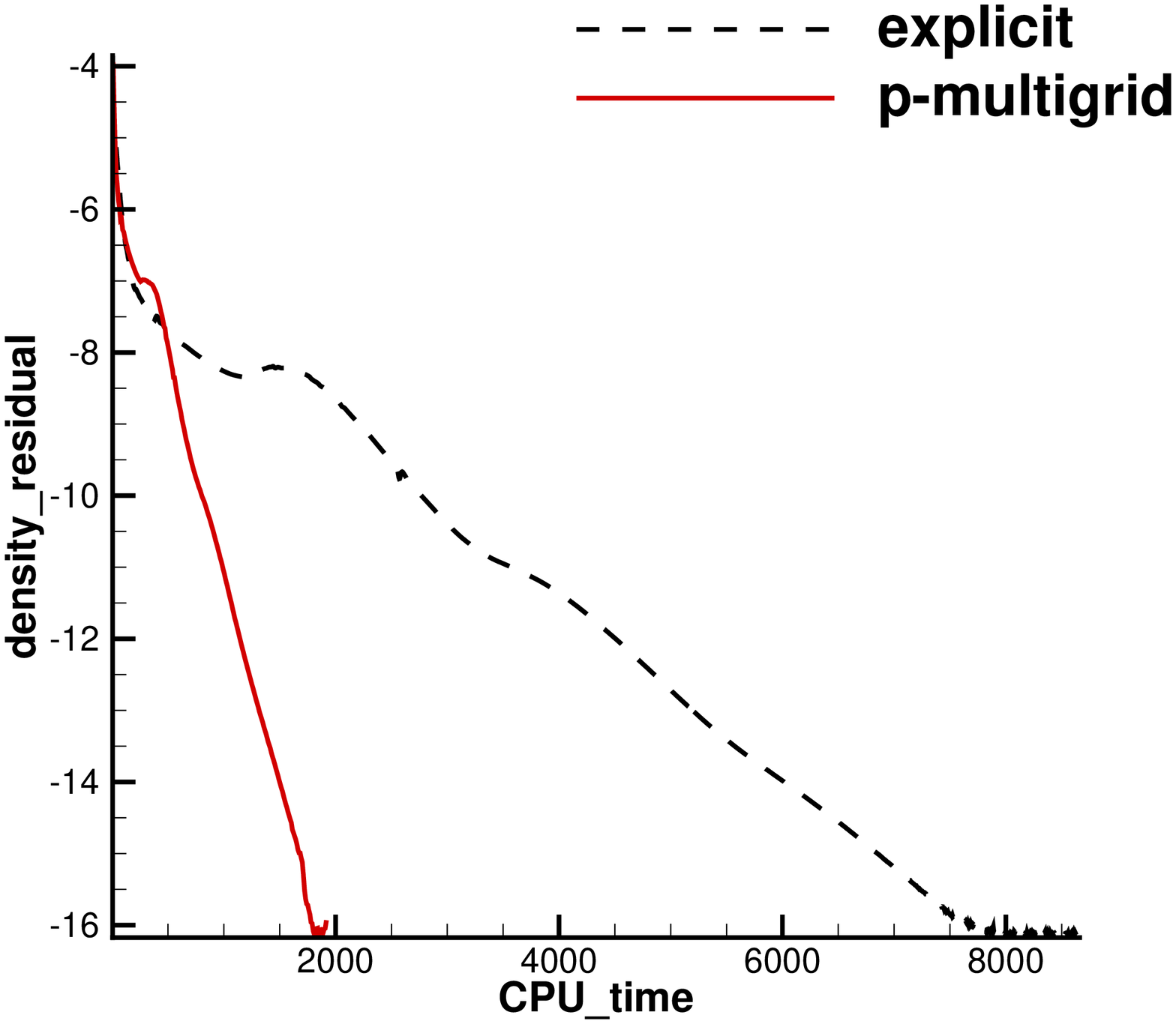}
	\includegraphics[width=0.32\textwidth]
    {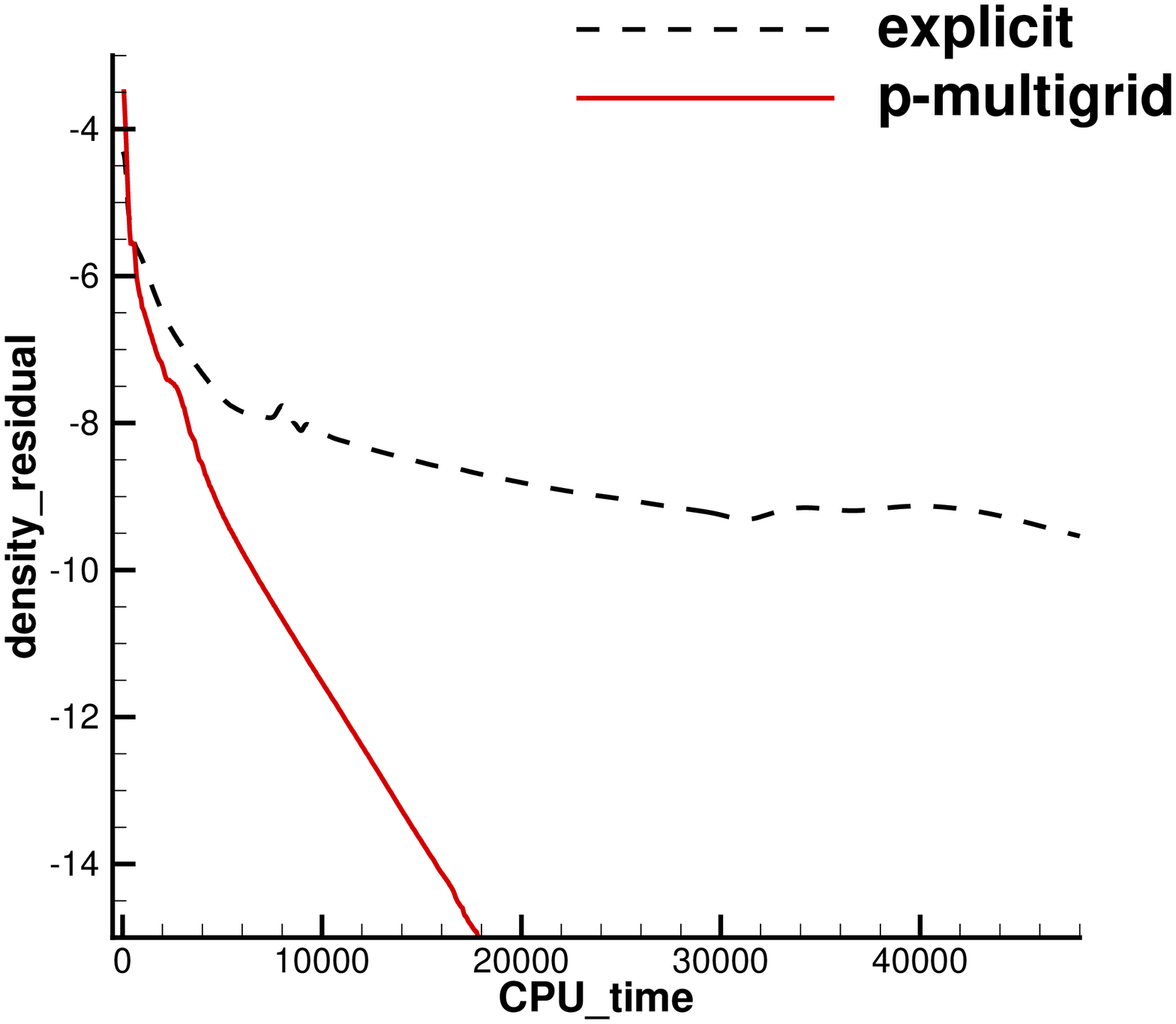}
	\vspace{-4mm} \caption{\label{cylinder-re40-res-cpu}
		Residual comparison between the explicit and p-multigrid CGKS.
		 Left: Mesh I. Right: Mesh II. }
\end{figure}

\begin{figure}[htp]	
	\centering
	\includegraphics[width=0.32\textwidth]
	{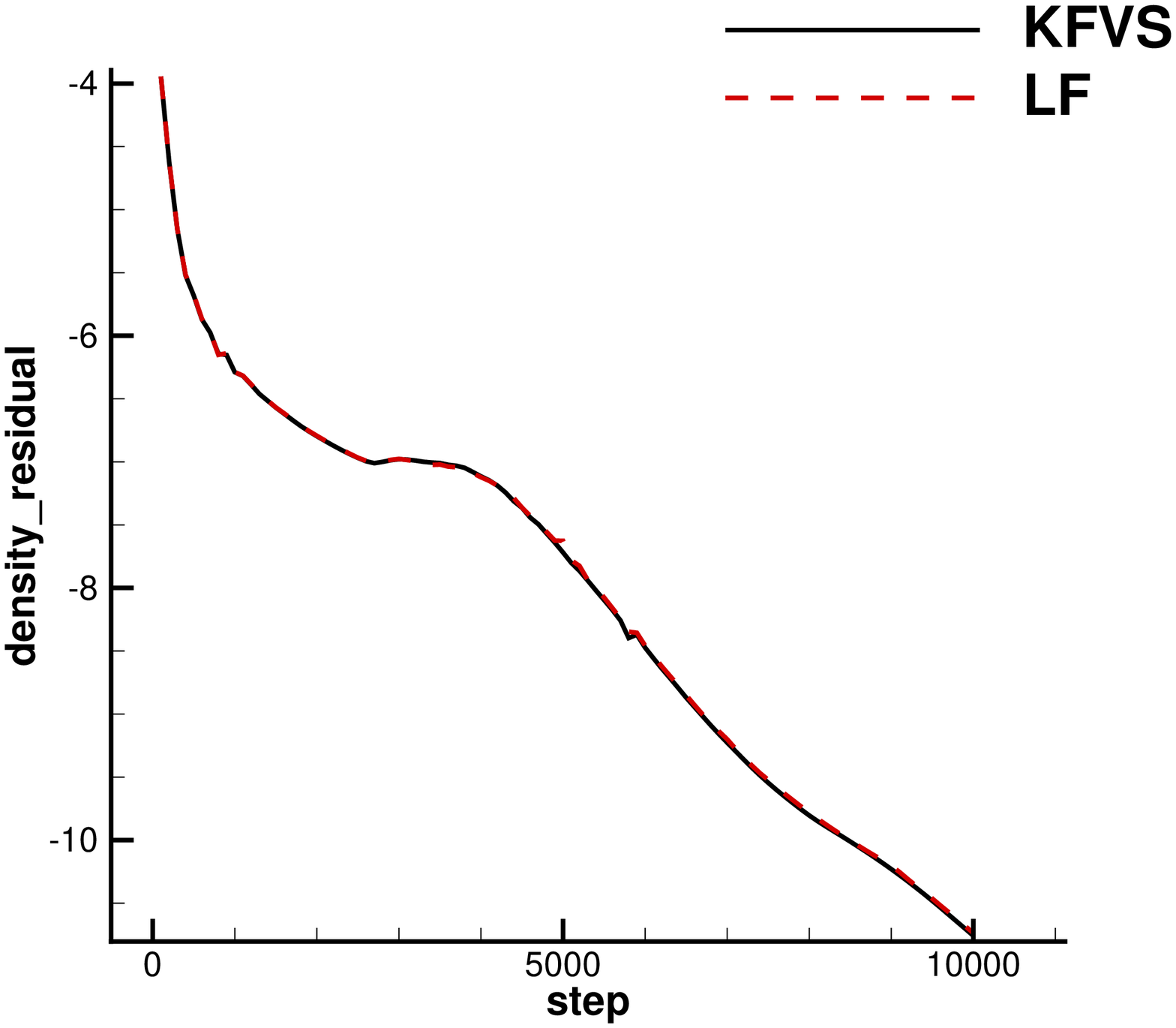}
	\includegraphics[width=0.32\textwidth]
{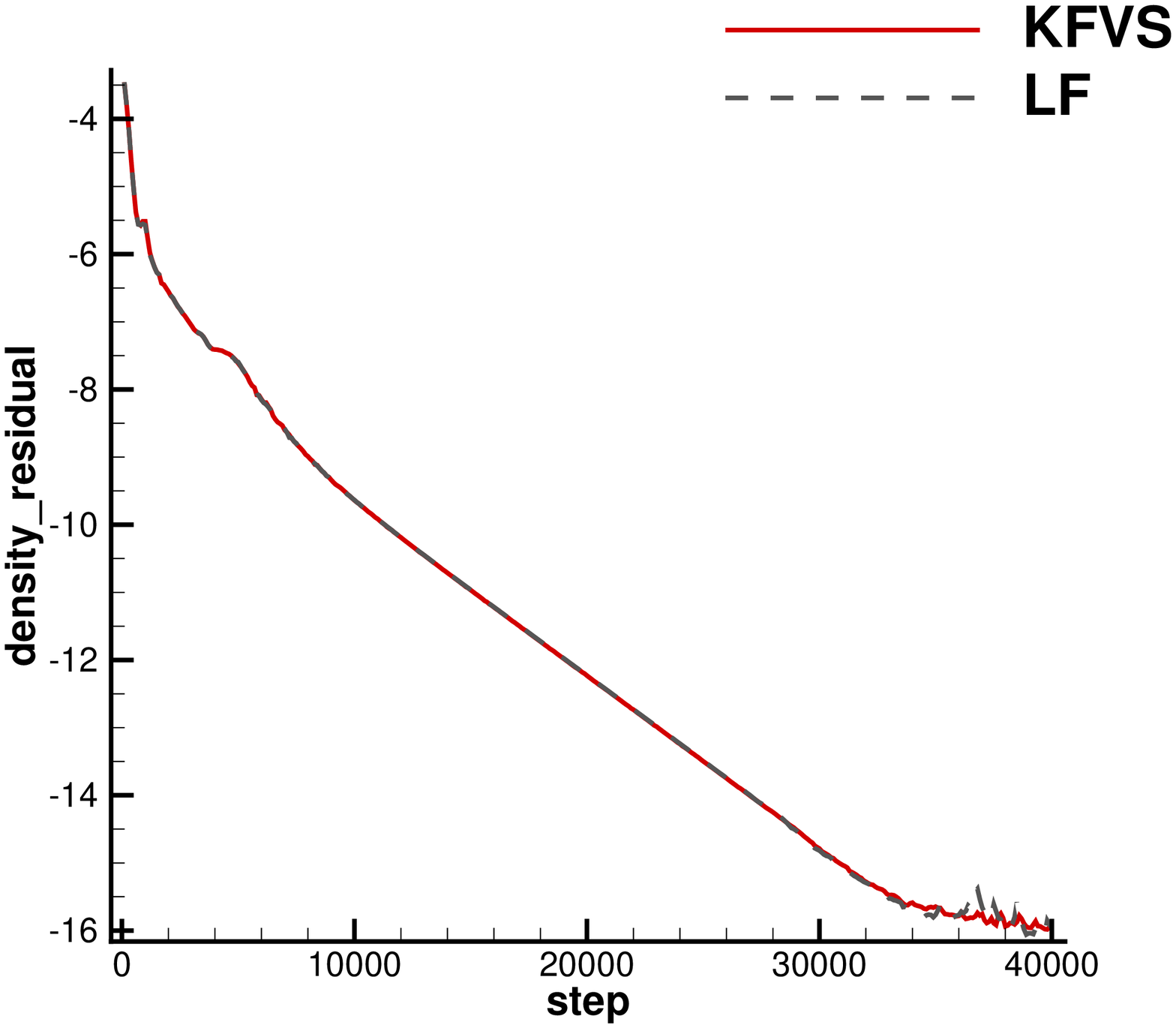}
	\vspace{-4mm} \caption{\label{cylinder-re40-res-flux}
The influence of different fluxes on the low-order level. Left: Mesh I. Right: Mesh II.  }
\end{figure}

\begin{figure}[htp]	
	\centering
	\includegraphics[width=0.32\textwidth]
{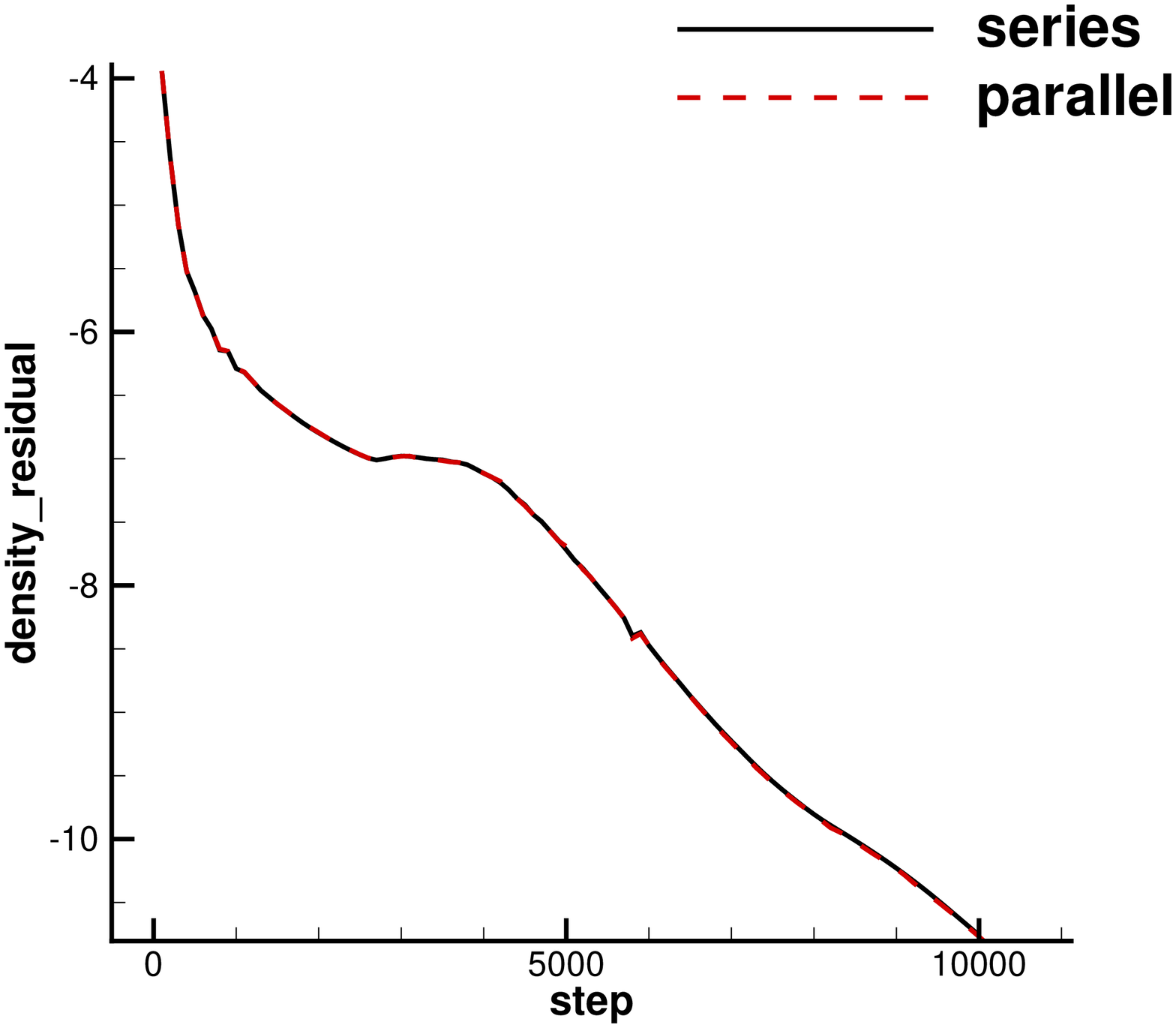}
	\includegraphics[width=0.32\textwidth]
	{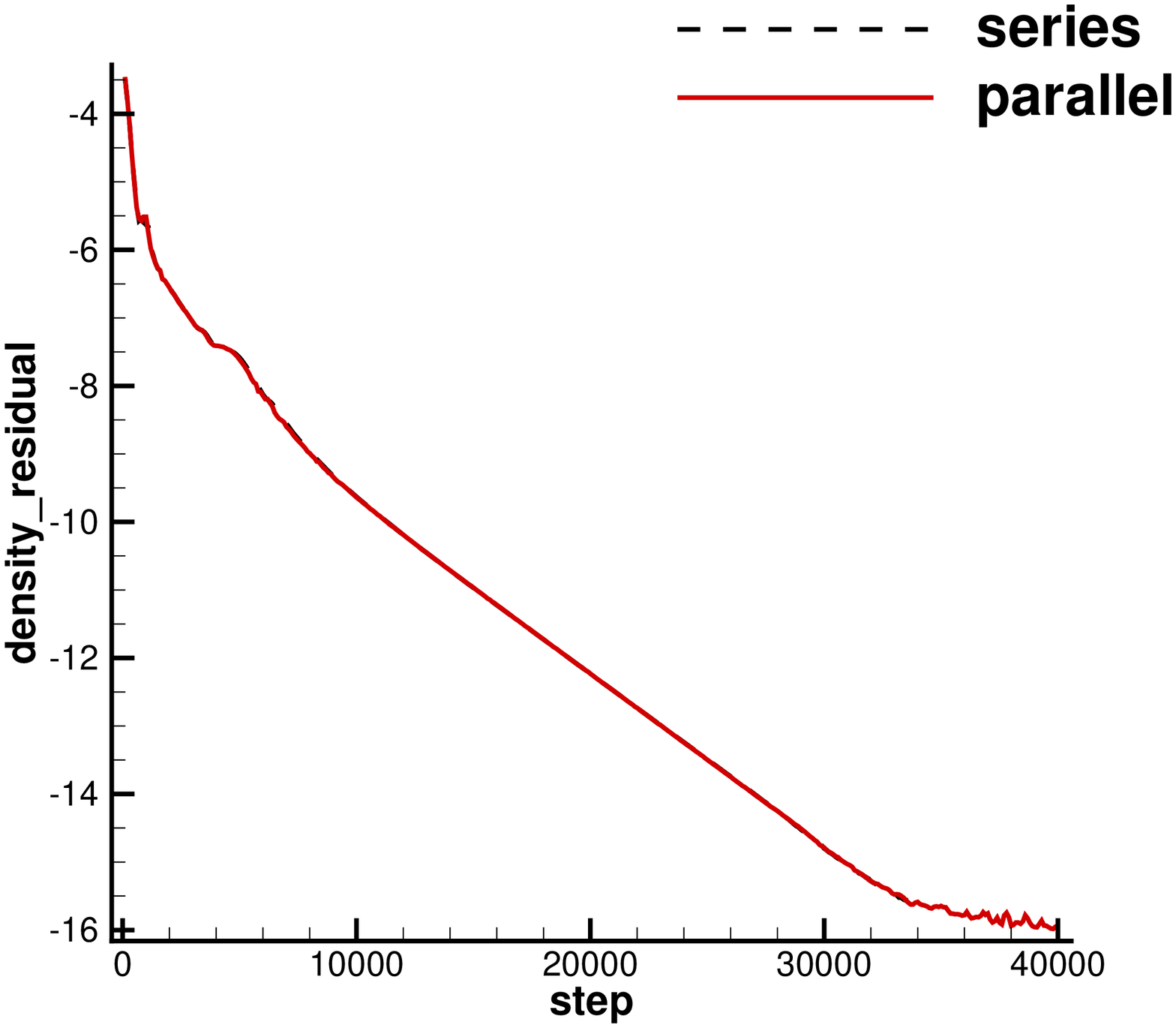}
	\vspace{-4mm} \caption{\label{cylinder-re40-res-para}
The influence of parallel computation on the low-order level. Left: Mesh I. Right: Mesh II.}
\end{figure}

A subsonic flow around a circular cylinder is simulated.
The Mach number is Ma=0.15 and Reynolds number Re=40 are based on the diameter of the cylinder $D=1$.
Two sets of mesh, namely Mesh I and Mesh II are used to evaluate the performance of the p-multigrid CGKS.
Mesh I is shown in Fig.~\ref{cylinder-re40-coarse-mesh} with a near wall size $h=1/20$.
A finer mesh, Mesh II is shown in in Fig.~\ref{cylinder-re40-fine-mesh} with a near wall size $h=1/96$.
A separation bubble is formed behind the cylinder, which is steady and symmetrical.
The CPU time history of the density residual is plotted in Fig.~\ref{cylinder-re40-res-cpu}.
The residuals for both explicit and p-multigrid CGKS on Mesh I settle down to machine zero and the speedup for the p-multigrid CGKS is about 4.
The residual for p-multigrid CGKS on Mesh II reduces to machine zero smoothly, while the explicit counterpart might need far more steps.
In this case, the speedup for the p-multigrid CGKS is about 8 for the same residual $10^{-9}$.
Then, the influence of different flux solvers used on the low-order level is investigated, as shown in Fig.~\ref{cylinder-re40-res-flux}.
Very close results are obtained for the KFVS solver and the Lax–Friedrichs (L-F) solver under both meshes.
It indicates that the usage of either kinetic-type or hydrodynamic-type solver on the low-order level has little effect on the performance of the proposed scheme.
The possible explanation is that the first-order scheme on the low-order level is much more dissipative than the CGKS on the high-order level, regardless of the choice of solvers.
Thirdly, the performance of the parallel computational efficiency is evaluated.
Eight threads are used for the parallel case.
It is exciting that the almost identical results are obtained for both series and parallel computation, as shown in Fig.~\ref{cylinder-re40-res-para}, which means no more iterations are required for parallel computation and the simple parallel strategy adopted here is successful.
The underlying reason might come to the sufficient number of sub-iterating steps on the low-order level.

\begin{figure}[htp]	
	\centering
	\includegraphics[width=0.4\textwidth]
	{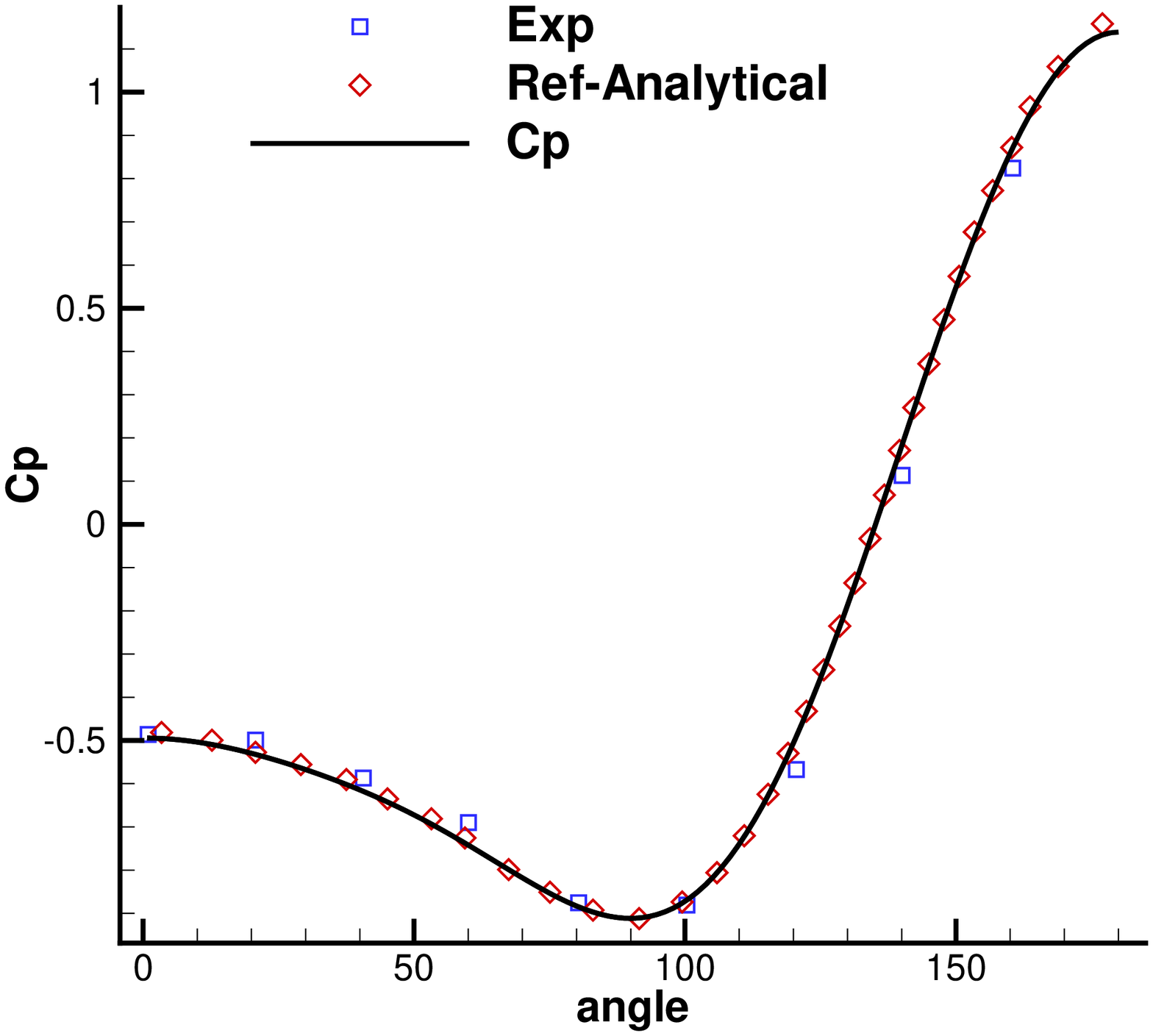}	
	\includegraphics[width=0.4\textwidth]
	{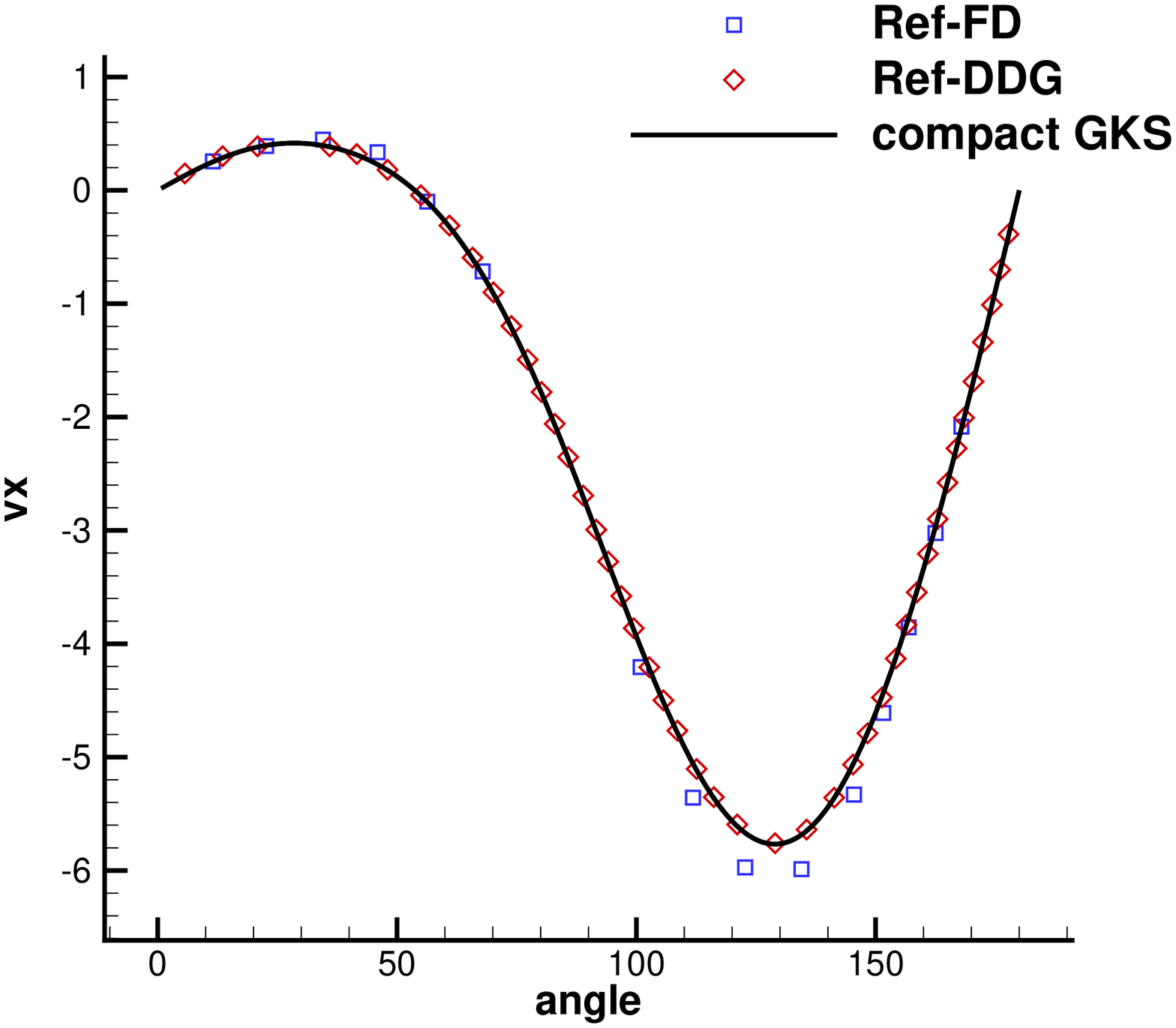}
	\vspace{-4mm} \caption{\label{cylinder-re40-line}
		Circular cylinder: Re=40. Left: surface pressure coefficient distribution. Right: surface local tangential velocity gradient  distribution. The results are obtained under Mesh II. }
\end{figure}

To validate the high resolution of the CGKS, the quantitative results under Mesh II including the drag and lift coefficients Cd,~Cl, the wake length $L$, and the separation angle $\theta$, etc are listed in Tab.~\ref{cylinder-re-40-cd-cl}, which agree well with the experimental and numerical references \cite{tritton1959experiments,coutanceau1977experimental,zhang2019direct}.
Furthermore, the quantities on the cylinder surface are extracted, including
the surface pressure coefficient Cp$=\frac{p-p_{\infty}}{\frac{1}{2}\rho_{\infty}U_{\infty}^2}$
and the non-dimensional local tangential velocity gradient$ \frac{2U_{\infty}}{D} \frac{\partial U_{\tau}}{\partial \eta}$, as shown in Fig.~\ref{cylinder-re40-line}.
The Cp from the current CGKS matches nicely with the experimental data \cite{grove1964experimental} and the analytical solution \cite{bharti2006steady}.
The tangential velocity gradient obtained by the current scheme is compared with those by the finite difference method \cite{braza1986Numerical} and the direct DG method \cite{zhang2019direct}.

\begin{table}[htp]
	\footnotesize
	\begin{center}
		\def\temptablewidth{1.0\textwidth}
		{\rule{\temptablewidth}{1pt}}
		\begin{tabular*}{\temptablewidth}{@{\extracolsep{\fill}}c|c|c|c|c|c|c}
			
			Case & Cd &Cl & Wake Length & Vortex Height & Vortex Width & Separation angle \\
			\hline
			Experiment \cite{tritton1959experiments} & 1.46 - 1.56 & -- & --   &  -- & --1 & --\\
			Experiment \cite{coutanceau1977experimental} & -- & -- & 2.12   &  0.297 & 0.751 & 53.5$^\circ$\\	
			DDG\cite{zhang2019direct} & 1.529 & -- & 2.31   &  -- & -- & --\\					
			current & 1.525 & 3.3e-14 & 2.22 & 0.296 & 0.714 & 53.3$^\circ$\\	
		\end{tabular*}
		{\rule{\temptablewidth}{0.1pt}}
	\end{center}
	\vspace{-4mm} \caption{\label{cylinder-re-40-cd-cl} Comparison of results for steady flow passing through a circular cylinder. The results are obtained under Mesh II.}
\end{table}

\subsection{Subsonic NACA0012 airfoil}

\noindent{\sl{(a) Inviscid case}}

\begin{figure}[htp]	
	\centering
	\includegraphics[width=0.9\textwidth]
	{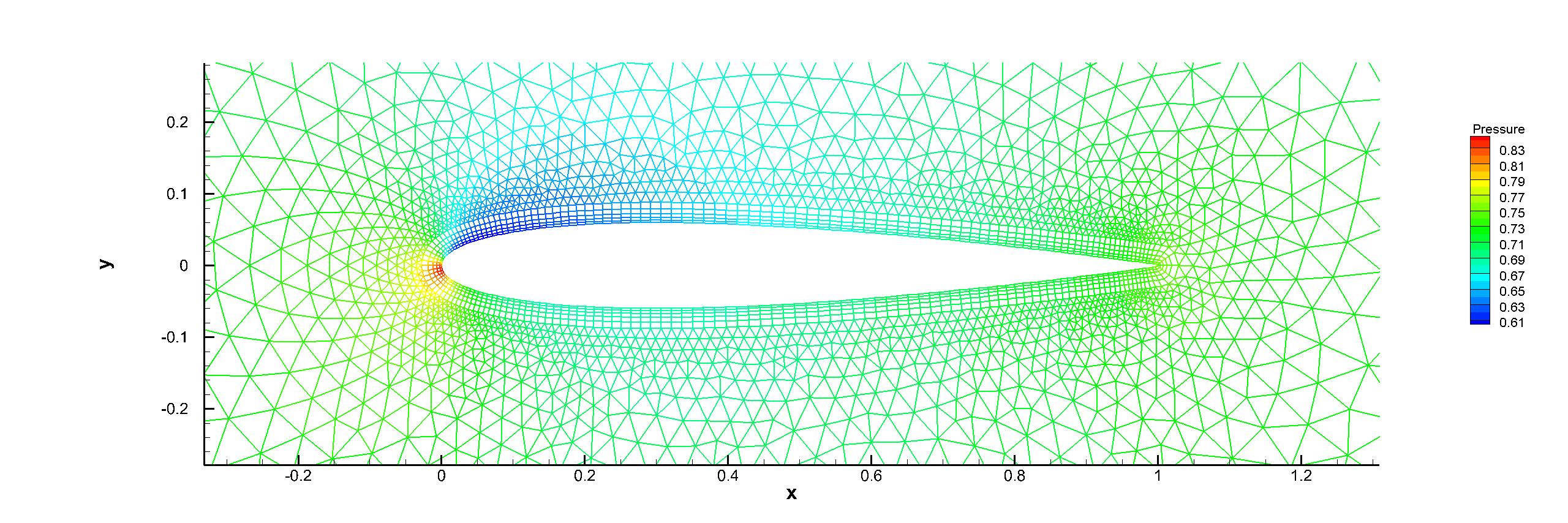}		
	\caption{Local mesh distribution colored by pressure for the inviscid NACA0012 airfoil case. Ma=0.5. AOA=2.0$^{\circ}$.}
	\label{naca0012-inviscid-mesh}
\end{figure}

\begin{figure}[htp]	
	\centering
	\includegraphics[width=0.32\textwidth]
	{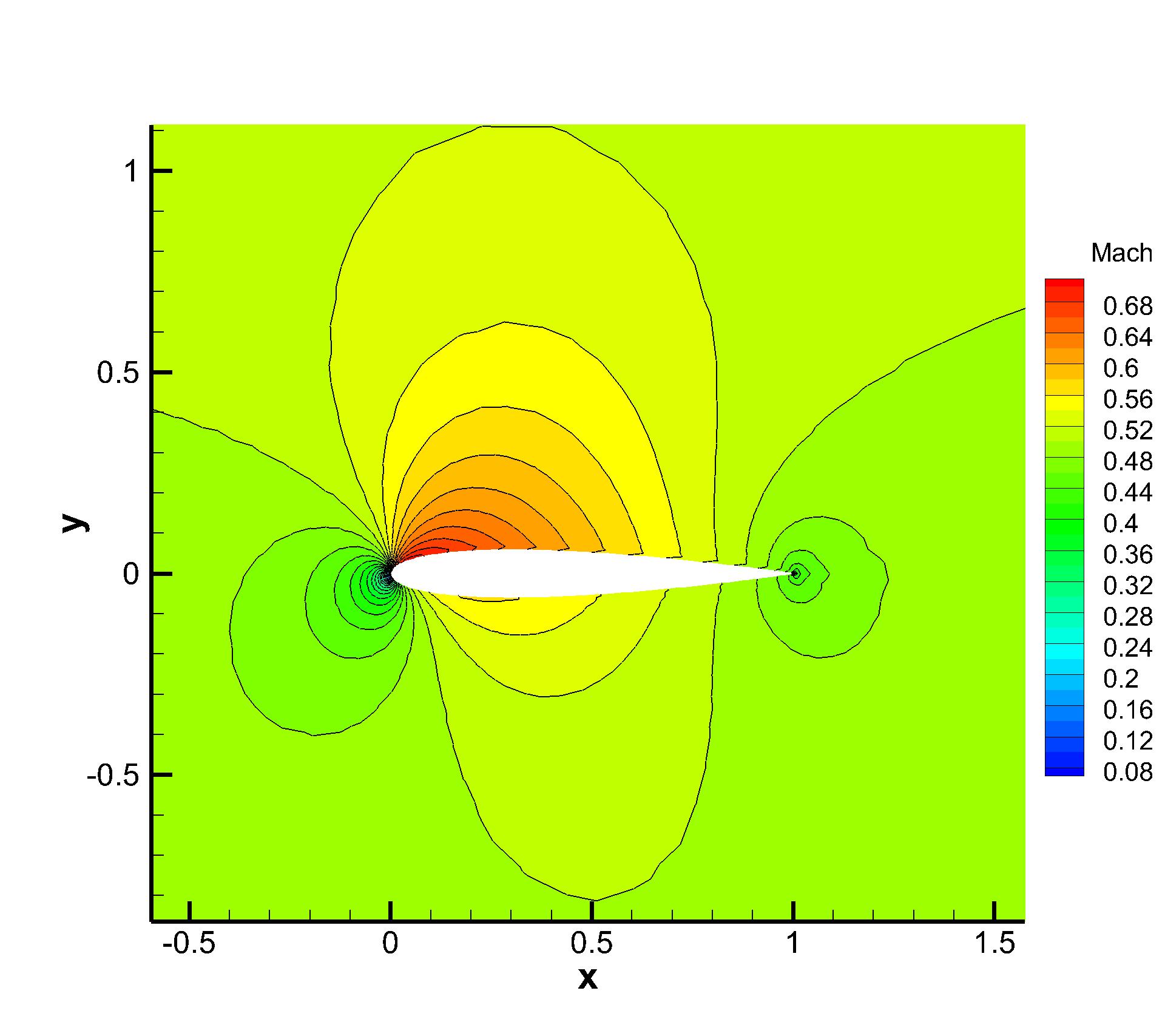}
	\includegraphics[width=0.32\textwidth]
	{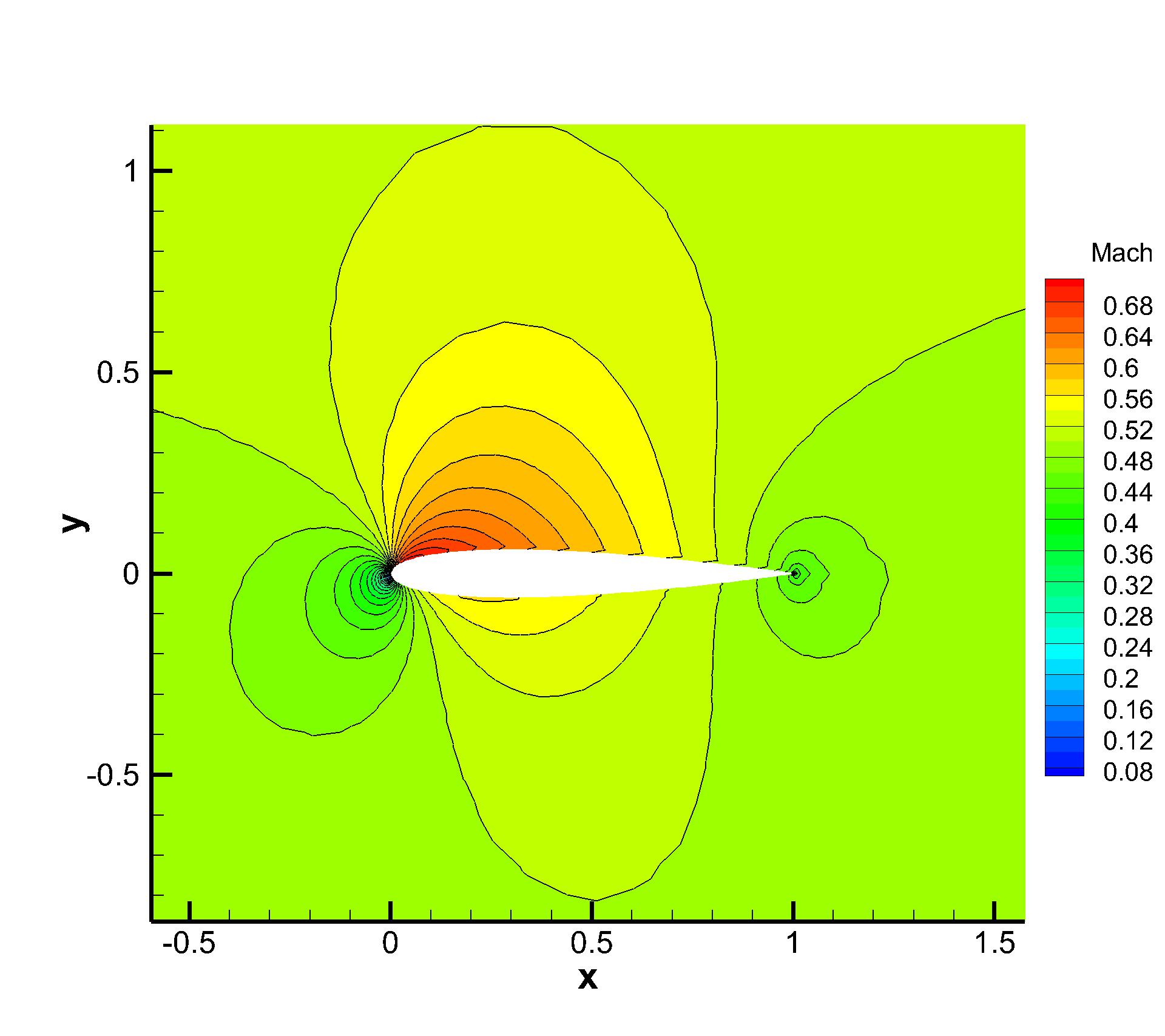}	
	\caption{Mach distributions for the inviscid NACA0012 airfoil case. Left: explicit  CGKS. Right: p-multigrid CGKS.}
	\label{naca0012-inviscid-ma}
\end{figure}

\begin{figure}[htp]	
	\centering
	\includegraphics[width=0.4\textwidth]
	{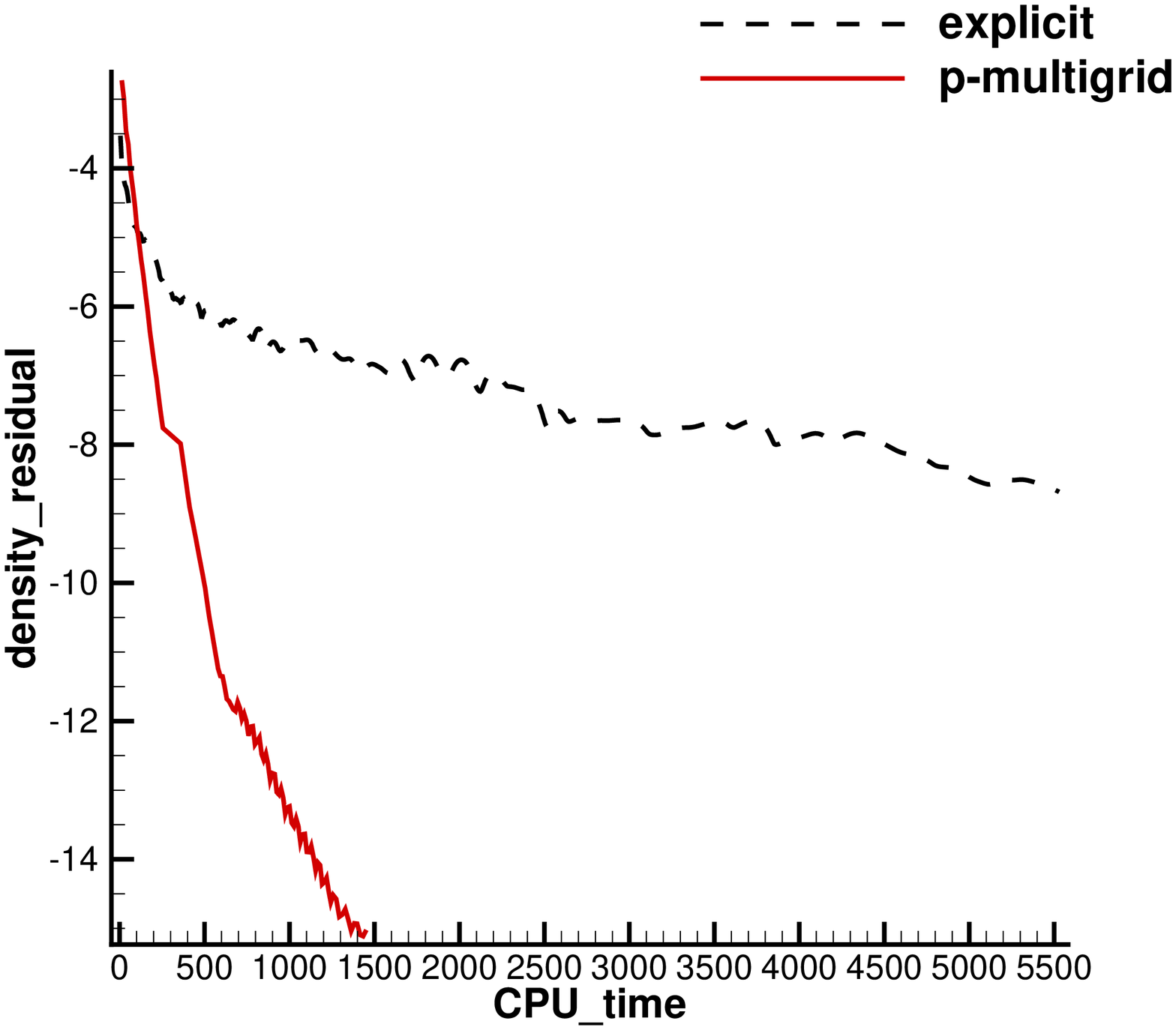}		
	\caption{Density residuals for the inviscid NACA0012 airfoil case.}
	\label{naca0012-inviscid-res}
\end{figure}

The inviscid flow passing through a NACA0012 airfoil with incoming Mach number Ma=0.5 and angle of attack (AOA) $\alpha=2.0^{\circ}$ is simulated first.
The reflective boundary condition is imposed on the airfoil surface
and the non-reflecting boundary condition based on the Riemann invariants is adopted on the far field.
Total $6538 \times 2$ hybrid prismatic cells are used in a cuboid domain $[-15,15]\times[15,15]\times[0,0.1]$.
The local unstructured mesh is shown in Fig.~\ref{naca0012-inviscid-mesh}, which is colored by the computed pressure distributions.
Almost identical Mach distributions are obtained for both explicit and p-multigrid CGKS, as given in Fig.~\ref{naca0012-inviscid-ma}.
The CPU time history of density residuals in Fig.~\ref{naca0012-inviscid-res} shows that the residual of p-multigrid CGKS can drop to $10^{-8}$ at around 350 seconds while the explicit CGKS needs about 4500 seconds.
A 13 times speedup is achieved in this case.

\noindent{\sl{(a) Viscous case}}

\begin{figure}[htp]	
	\centering
	\includegraphics[height=0.25\textwidth]
	{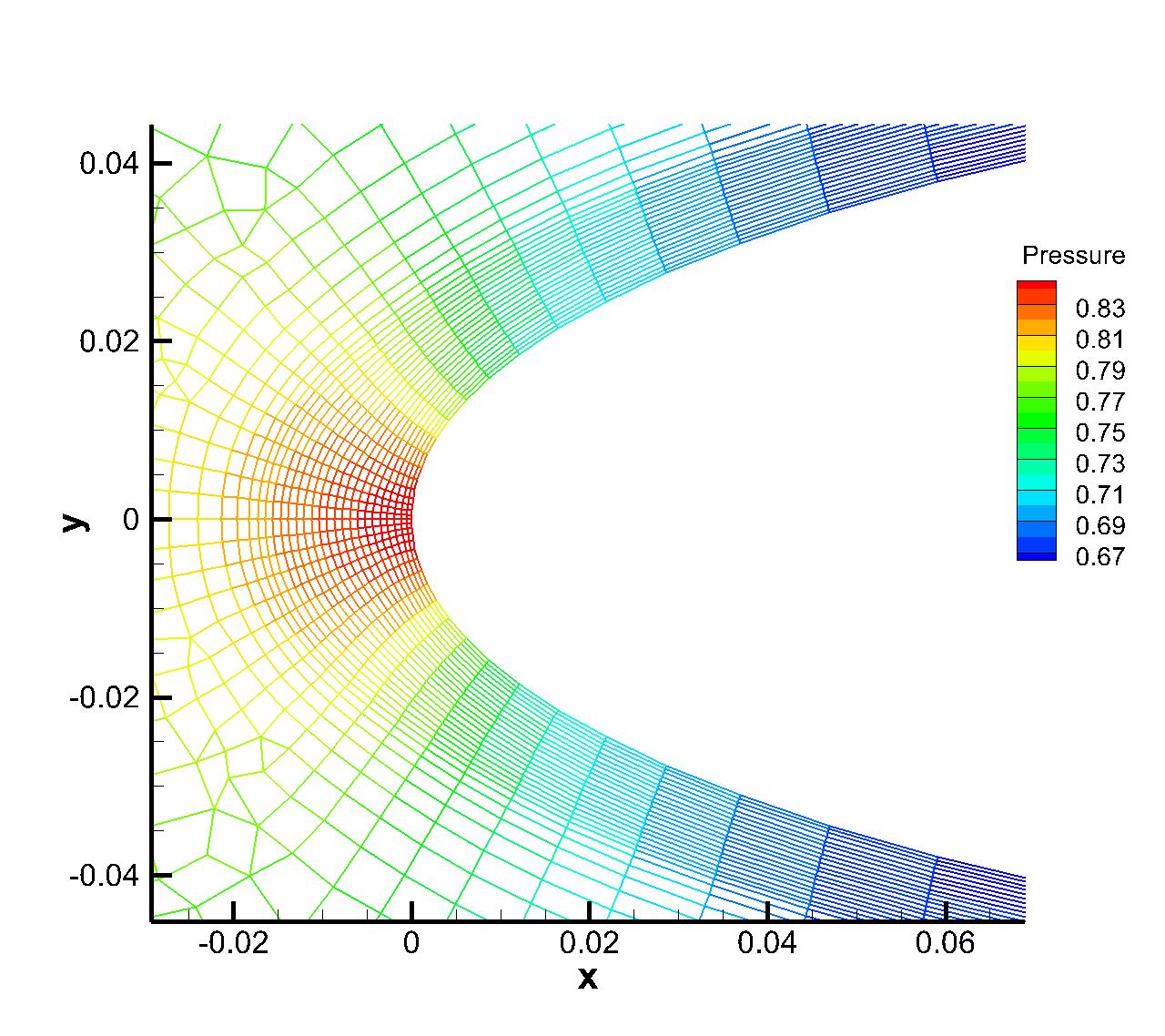}
		\includegraphics[height=0.25\textwidth]
	{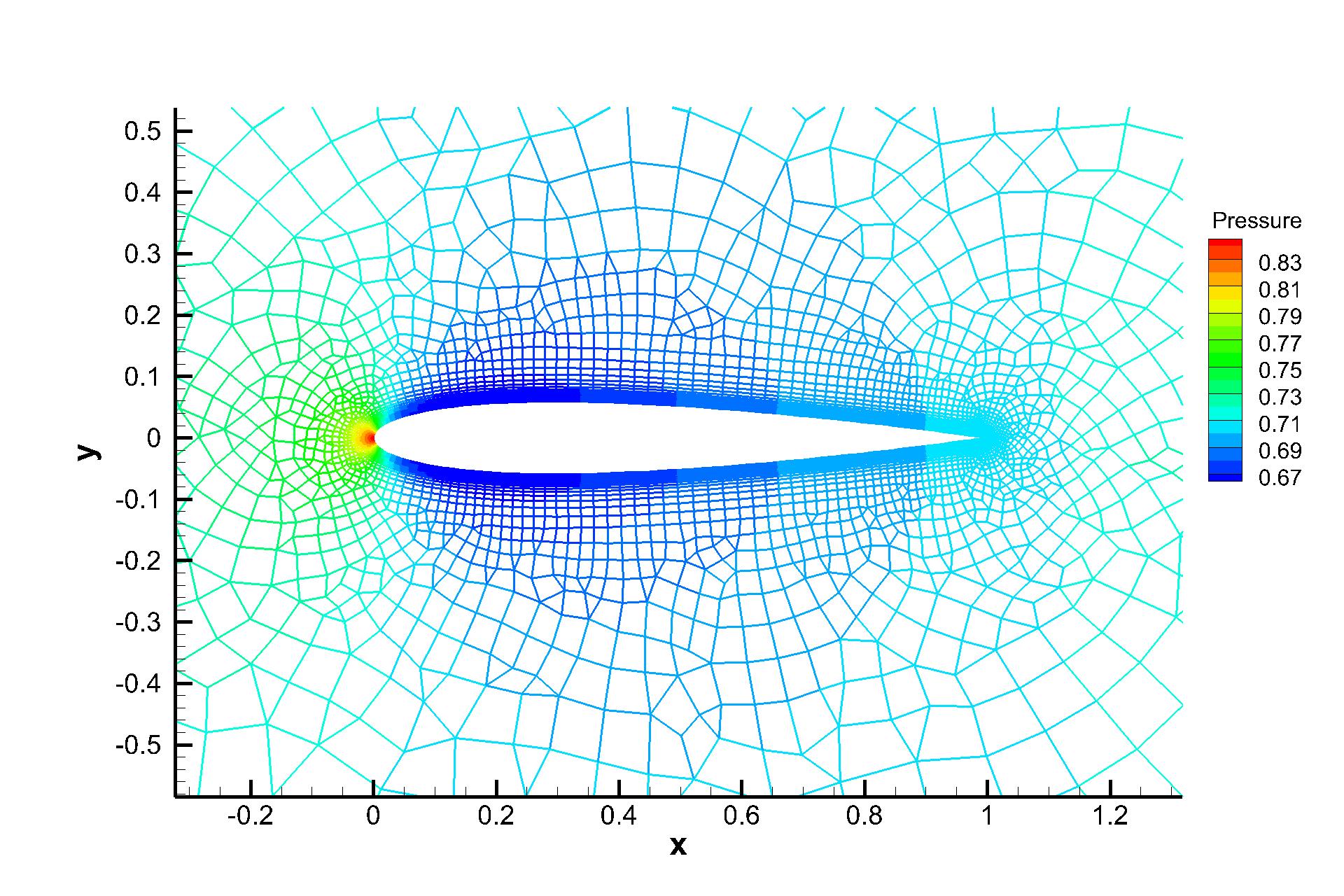}
		\includegraphics[height=0.25\textwidth]
	{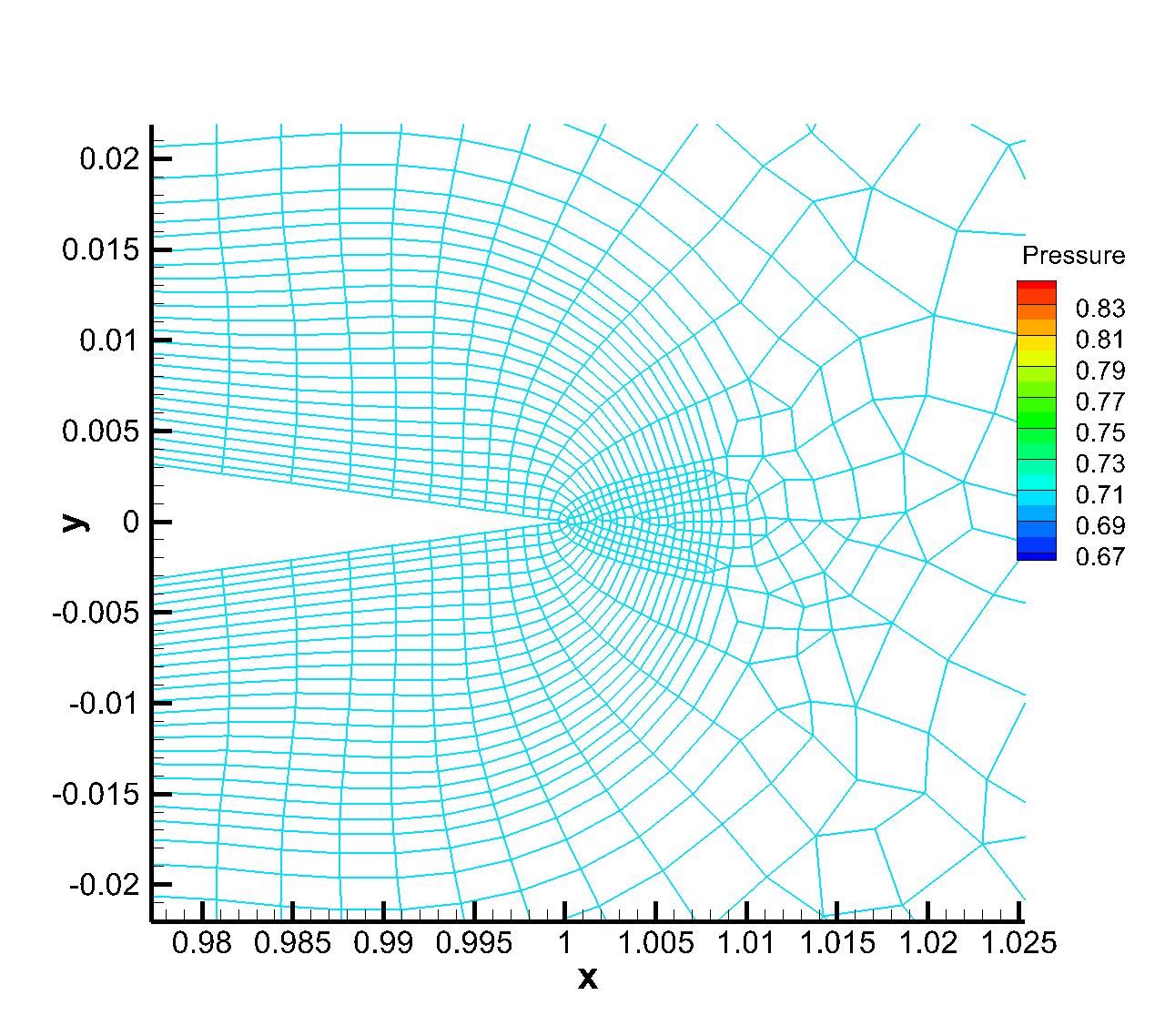}	
	\caption{Local mesh distributions colored by pressure for the viscous NACA0012 airfoil case. Ma=0.5. Re=500. AOA=0$^{\circ}$.}
	\label{naca0012-viscous-mesh}
\end{figure}

\begin{figure}[htp]	
	\centering
	\includegraphics[width=0.4\textwidth]
	{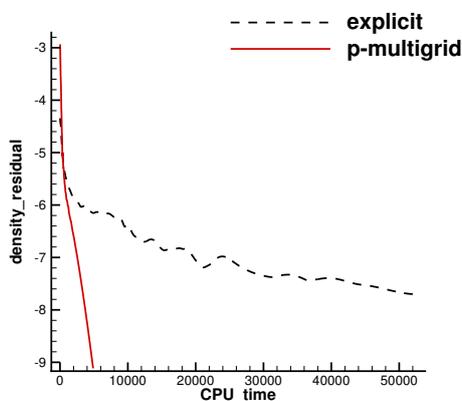}		
	\caption{Density residuals for the viscous NACA0012 airfoil case.}
	\label{naca0012-viscous-res}
\end{figure}

\begin{figure}[htp]	
	\centering
	\includegraphics[width=0.55\textwidth]
	{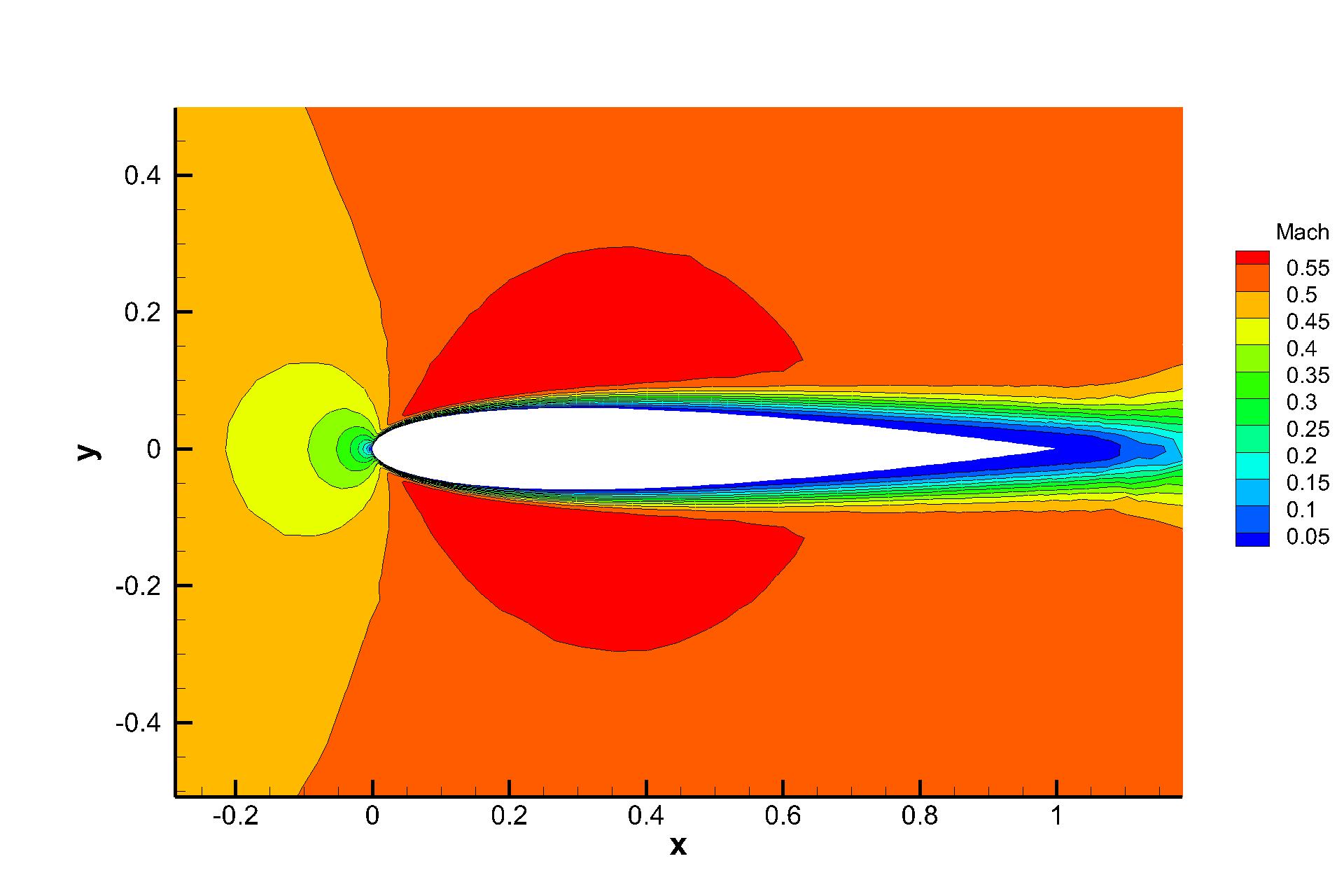}		
	\caption{Mach distributions for the viscous NACA0012 airfoil case obtained by the p-multigrid CGKS.}
	\label{naca0012-viscous-ma}
\end{figure}

\begin{figure}[htp]	
	\centering
	\includegraphics[width=0.4\textwidth]
	{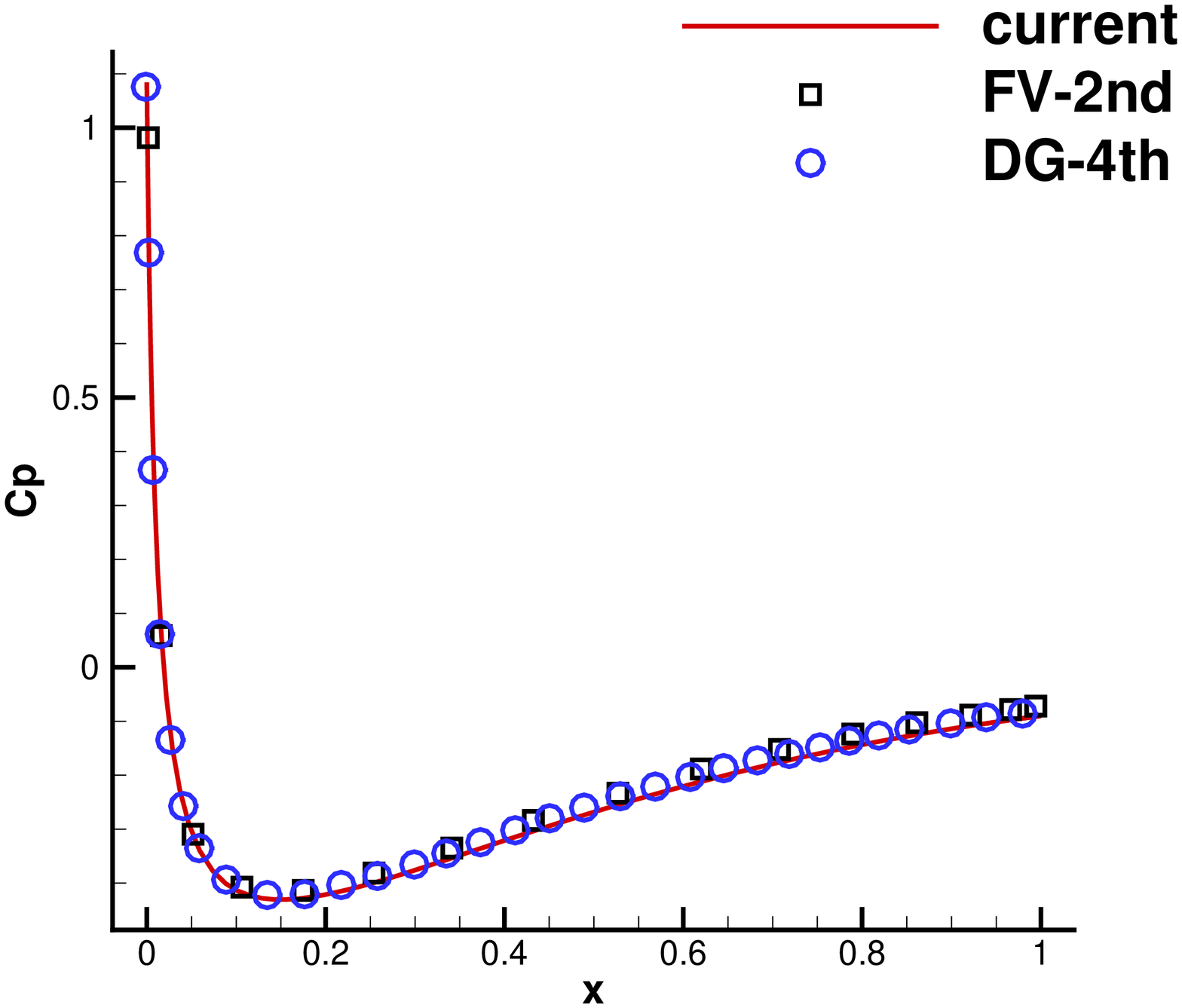}
		\includegraphics[width=0.4\textwidth]
	{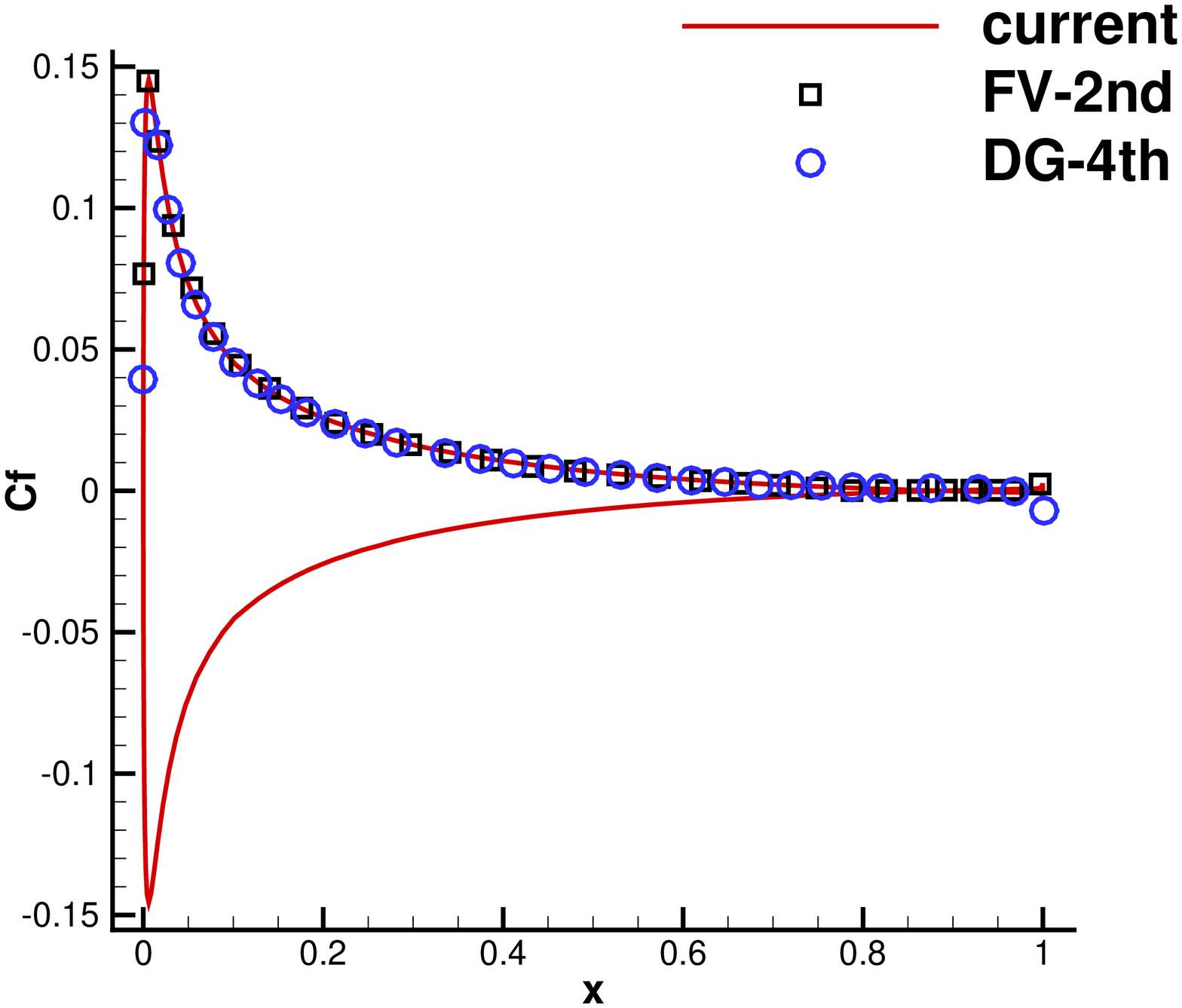}			
	\caption{Surface pressure coefficient (left) and skin friction coefficient (right) for the viscous NACA0012 airfoil case obtained by the p-multigrid CGKS.}
	\label{naca0012-viscous-line}
\end{figure}

A viscous flow passing through a NACA0012 airfoil is simulated then.
The Reynolds number is Re=5000 based on the chord length $L=1$. The incoming Mach number is Ma=0.5 and the AOA is 0$^{\circ}$.
The non-slip adiabatic boundary condition is imposed on the airfoil surface.
Unstructured mixed-element mesh with $7922 \times 2$ cells is used with a near wall size $h=4\times10^{-4}$. The maximum aspect ratio of cells is about 62.
The local enlargements of the mesh are presented in Fig.~\ref{naca0012-viscous-mesh}.
For this case, the speedup is about 15 at a residual level of $10^{-8}$, as shown in Fig.~\ref{naca0012-viscous-res}.
Computed pressure and Mach distributions obtained by the current scheme are shown in Fig.~\ref{naca0012-viscous-mesh} and Fig.~\ref{naca0012-viscous-ma}.
Quantitative results including the surface pressure coefficient and skin friction coefficient are extracted and plotted in Fig.~\ref{naca0012-viscous-line}, which agree nicely with the reference data from the second-order finite-volume method \cite{jawahar2000high} and fourth-order DG method \cite{bassi1997ns}.

\subsection{Shock reflection problem}
\begin{figure}[htp]	
	\centering
	\includegraphics[width=0.4\textwidth]
	{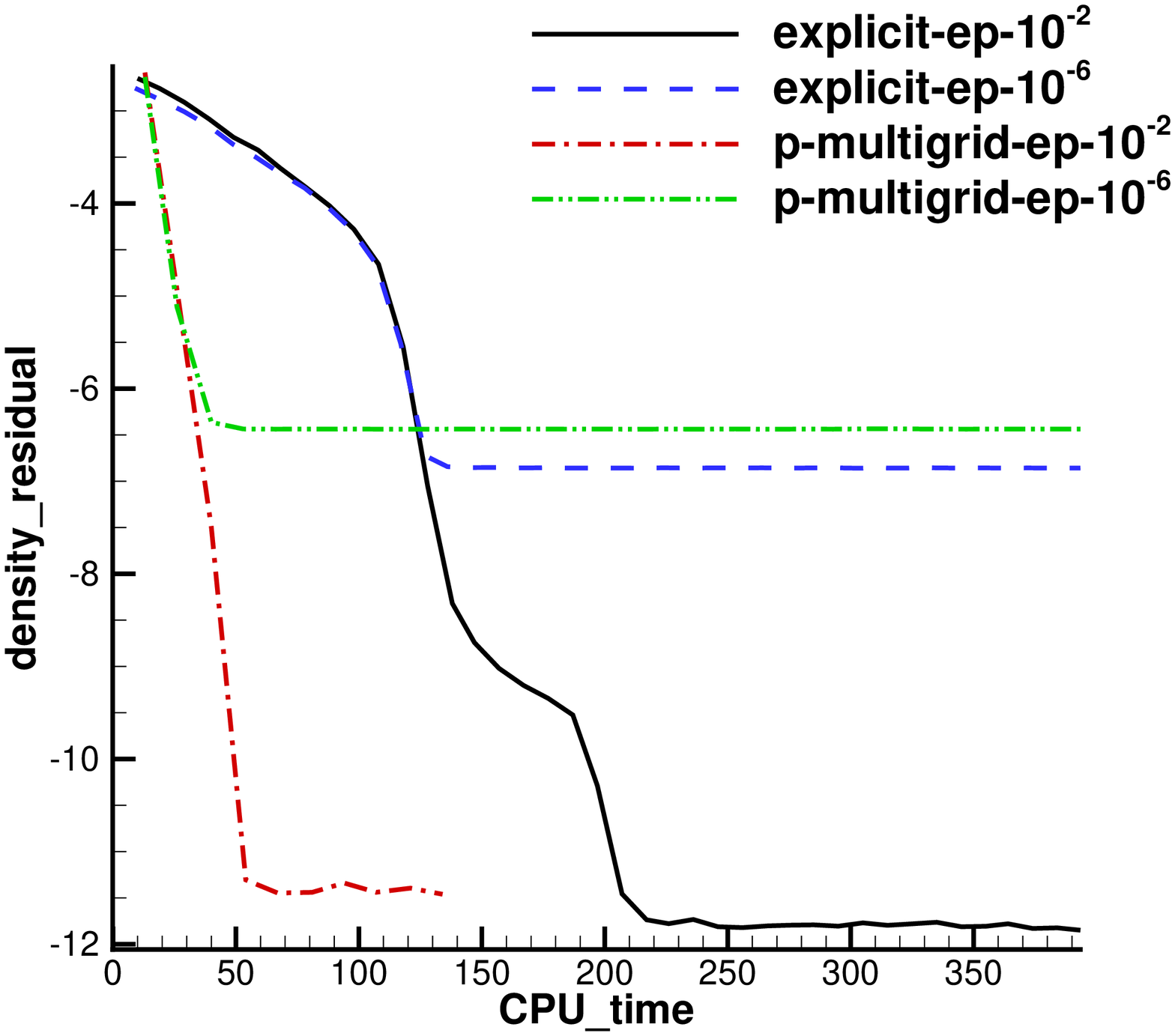}	
	\caption{Density residuals for the shock reflection problem. }
	\label{shock-reflection-res}
\end{figure}

\begin{figure}[htp]	
	\centering
	\includegraphics[width=0.8\textwidth]
	{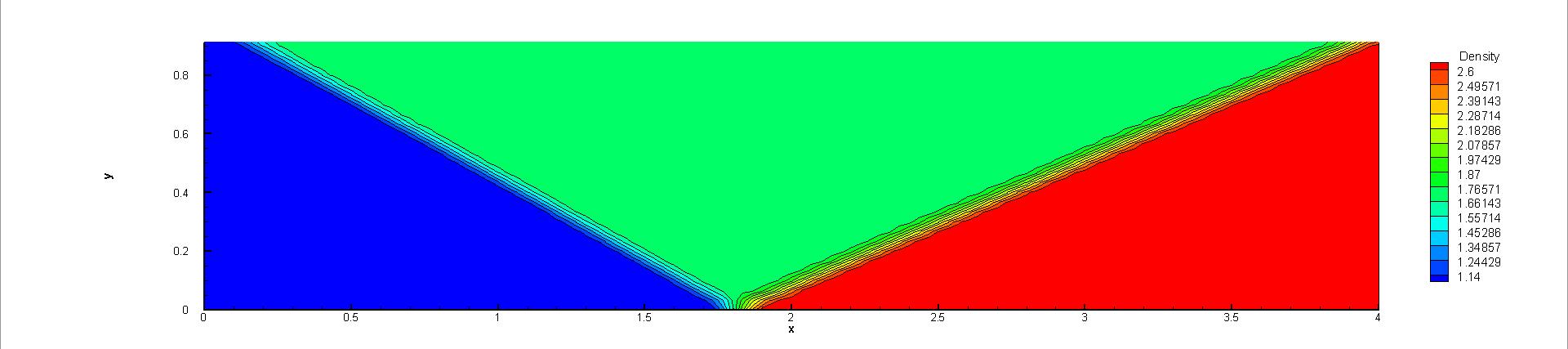}
	\includegraphics[width=0.8\textwidth]
	{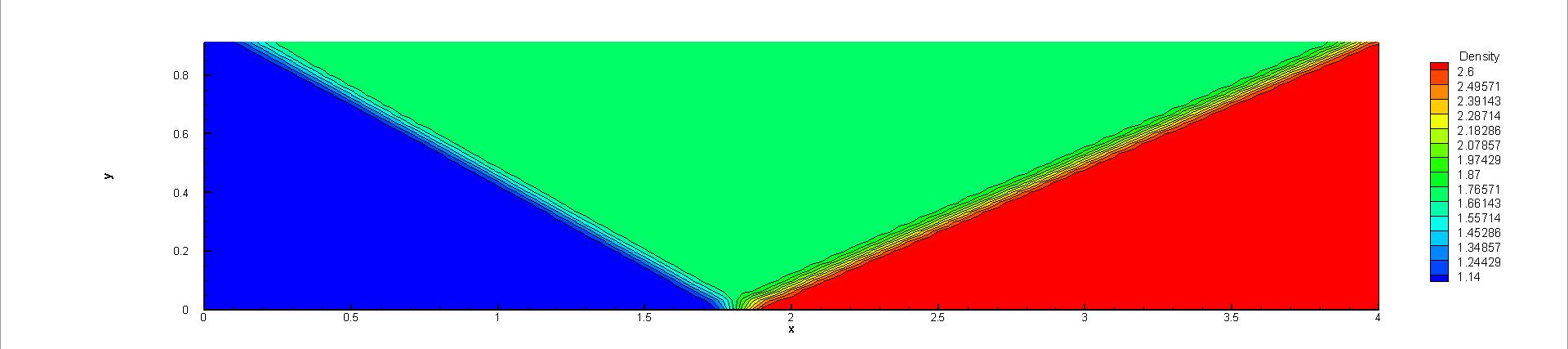}	
	\caption{Shock reflection problem. Up:explicit CGKS. Down:p-multigrid CGKS. $\epsilon=10^{-2}$. }
	\label{shock-reflection-ep-1e-2}
\end{figure}

\begin{figure}[htp]	
	\centering
	\includegraphics[width=0.8\textwidth]
	{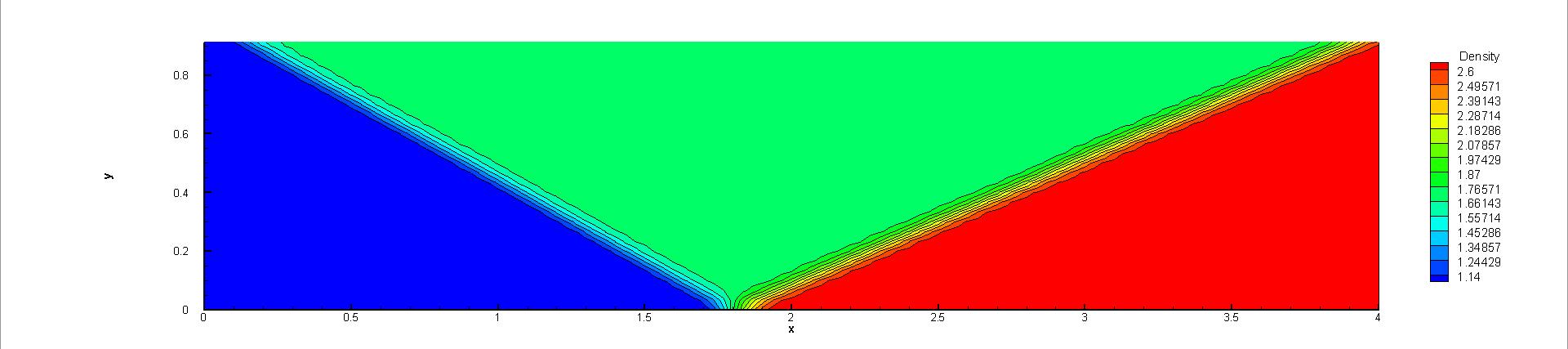}
	\includegraphics[width=0.8\textwidth]
	{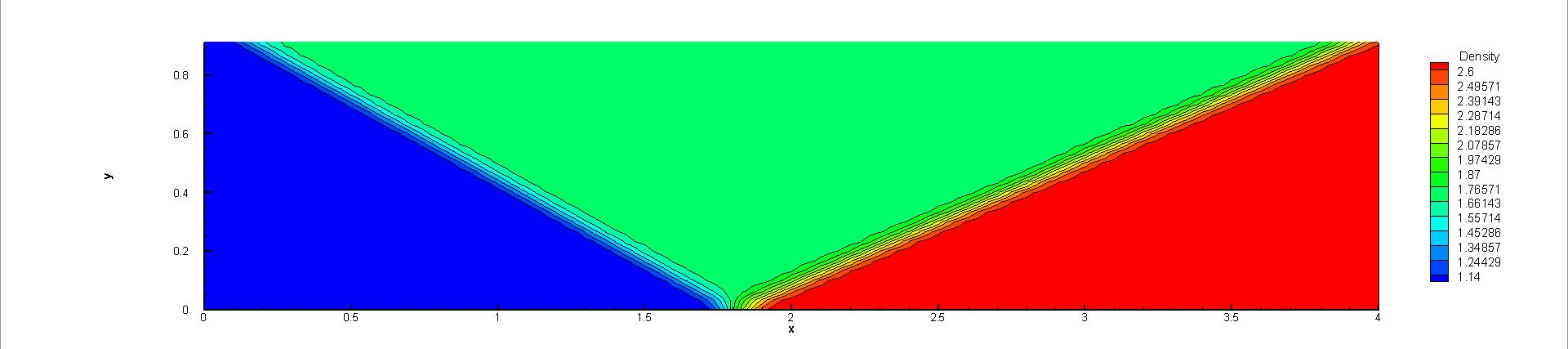}	
	\caption{Shock reflection problem. Up:explicit CGKS. Down:p-multigrid CGKS. $\epsilon=10^{-6}$. }
	\label{shock-reflection-ep-1e-6}
\end{figure}

\begin{figure}[htp]	
	\centering
	\includegraphics[width=0.32\textwidth]
	{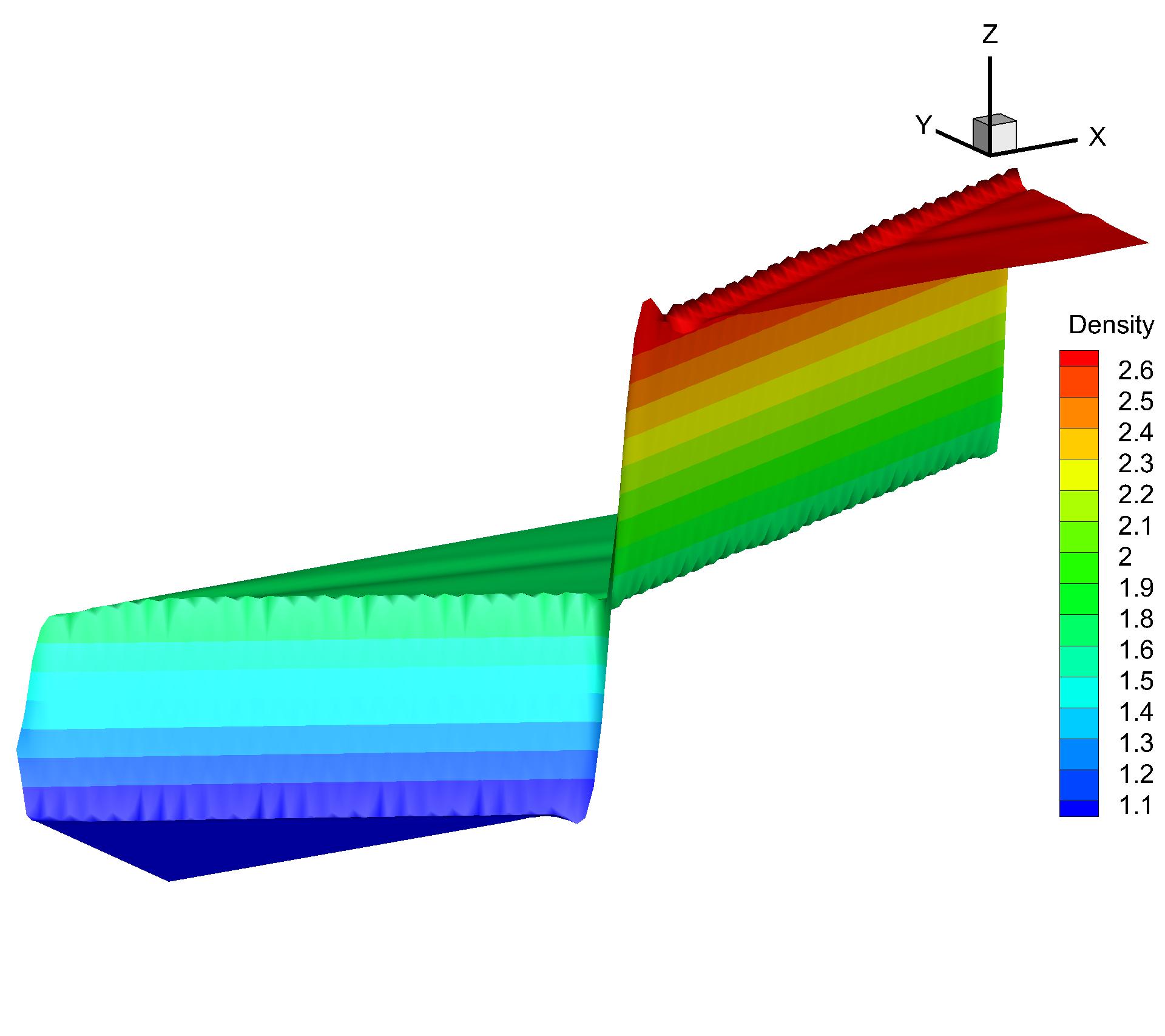}
	\includegraphics[width=0.32\textwidth]
	{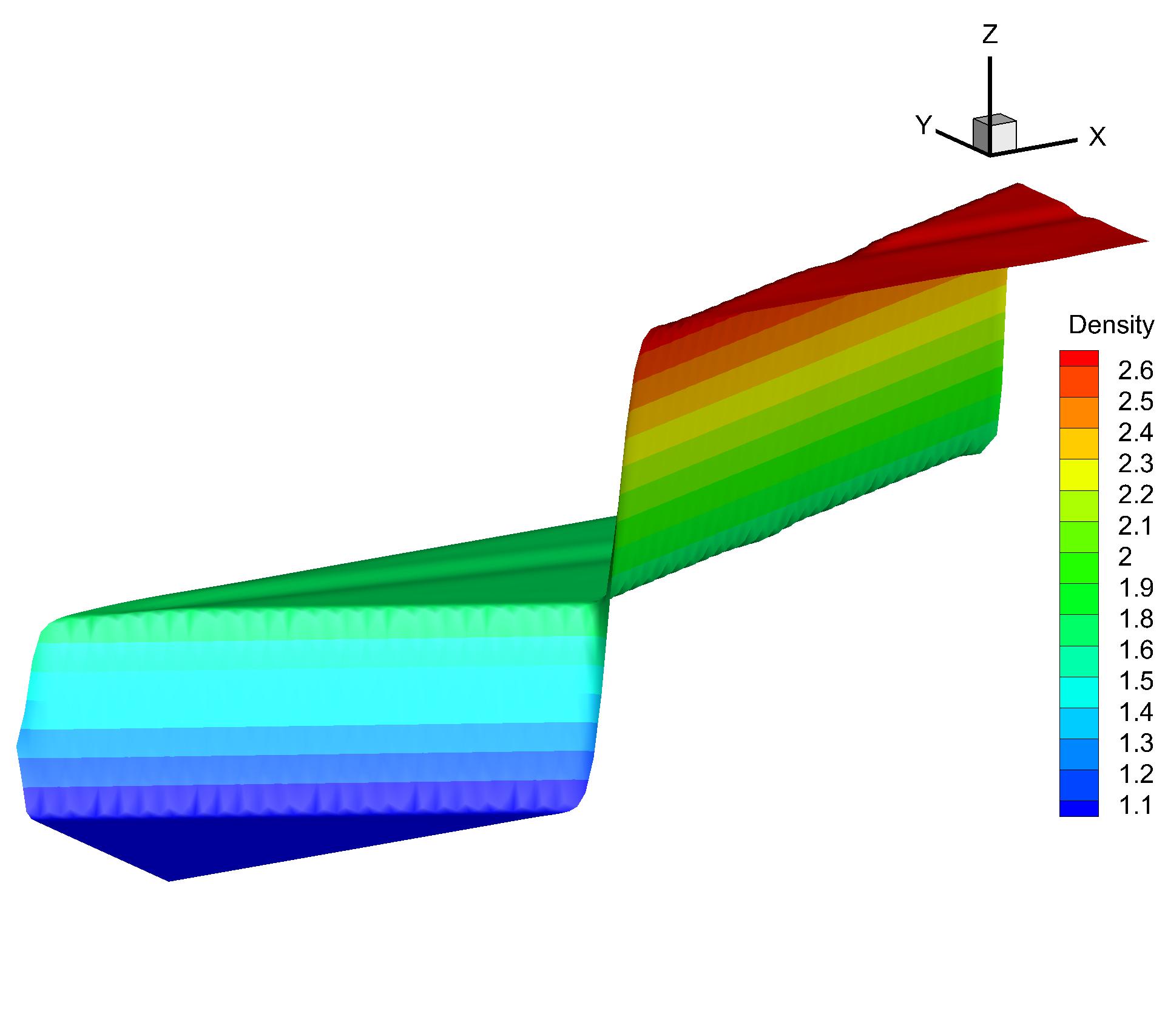}
	\caption{3-D view of the density distributions of the shock reflection problem obtained by the p-multigrid CGKS. Left: $\epsilon=10^{-2}$.
		Right: $\epsilon=10^{-6}$. }
	\label{shock-reflection-3d}
\end{figure}

It is reported that the high-order methods can easily fail to be convergent for flow with shocks \cite{zhang2019brief}.
The shock reflection problem is tested to validate the performance of CGKS for solving steady-state problems containing shocks.
The geometry is a rectangle with length 4 and height 1.
A Cartesian mesh with $120 \times 30 \times 2$ cells is used.
The reflective wall is given on the bottom of the computational domain.
The supersonic outflow boundary condition  is imposed on the right.
The other two sides are Dirichlet conditions:
\begin{align*}
(\rho,U_1,U_2, p)&=(1.0, 2.9, 0,
1.0/1.4)|_{\text{left}},\\
(\rho,U_1,U_2, p)&=(1.69997, 2.61934, -0.50632, 1.52819)|_{\text{up}}.
\end{align*}
A uniform flow with the value of left boundary is set initially.
To avoid undesired spurious oscillations around discontinuities,
the reconstruction based on the characteristic variables is adopted.
Since the numerical convergence of this problem is sensitive to the value of small positive number $\epsilon$ in the compact WENO reconstruction \cite{ji2021gradient}, two $\epsilon$, i.e., $10^{-2}$ and $10^{-6}$ are chosen in the test.
Through the CPU time history of the density residuals in Fig.~\ref{shock-reflection-res},
it can be observed that with $\epsilon=10^{-2}$, both explicit and p-multigrid CGKS can convergence nicely with a residual level less than $10^{-11}$.
However, with $\epsilon=10^{-6}$, the errors for both schemes settle around $10^{-7}$.
About 3 times of speedup is achieved in this case.
On the other hand, there is barely any difference that can be observed in  2-D density contours extracted from the central $x-y$ plane with the same $\epsilon$,
as shown in Fig.~\ref{shock-reflection-ep-1e-2} and Fig.~\ref{shock-reflection-ep-1e-6}.
In addition, with the increasing of $\epsilon$, the non-linear weights are closer to the linear weights and the numerical solutions near the shock become slightly oscillatory, as shown in Fig.~\ref{shock-reflection-3d}.

\subsection{Transonic dual NACA0012 airfoils}

\begin{figure}[htp]	
	\centering	
	\includegraphics[width=0.4\textwidth]
	{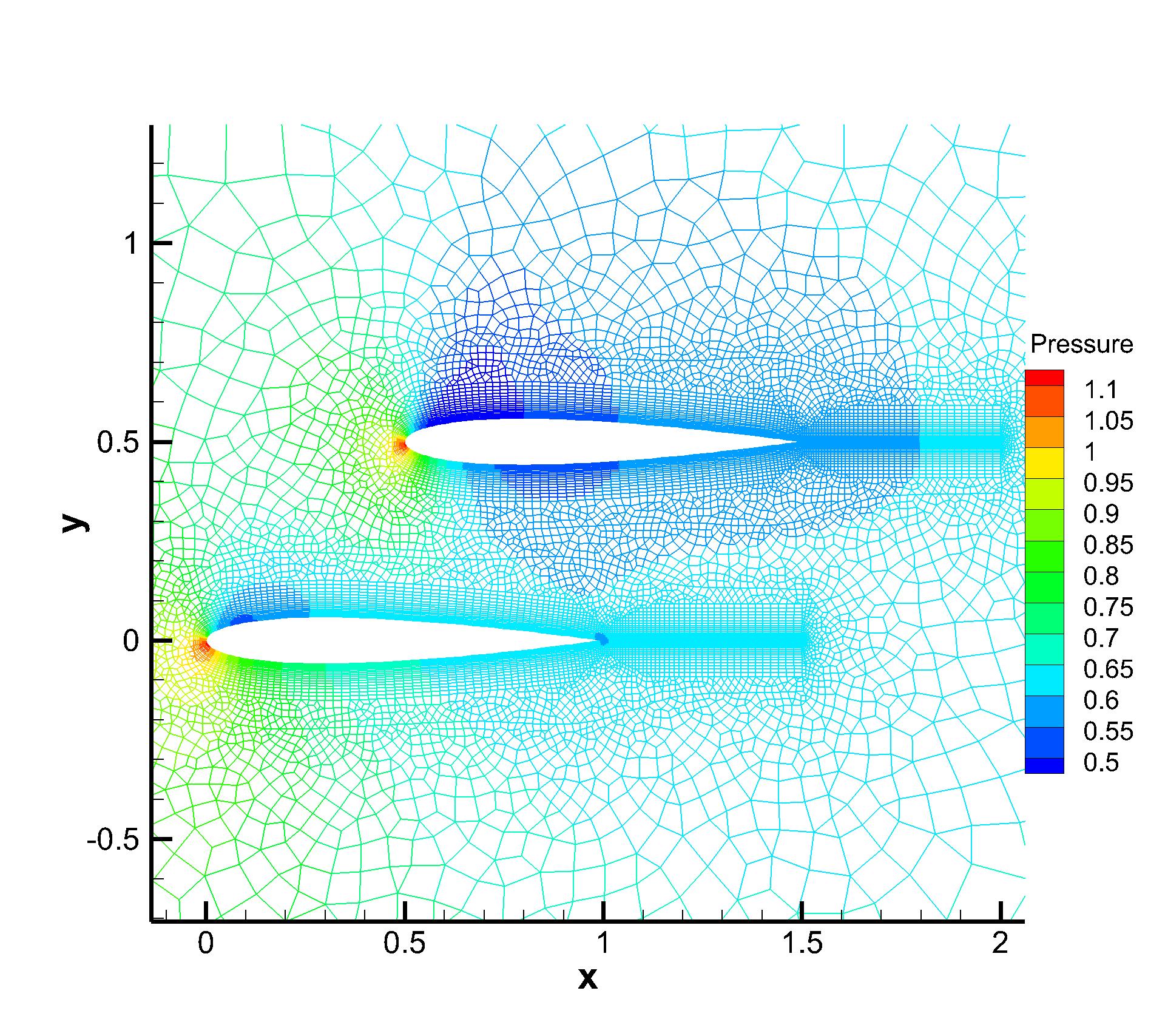}
	\includegraphics[width=0.4\textwidth]
	{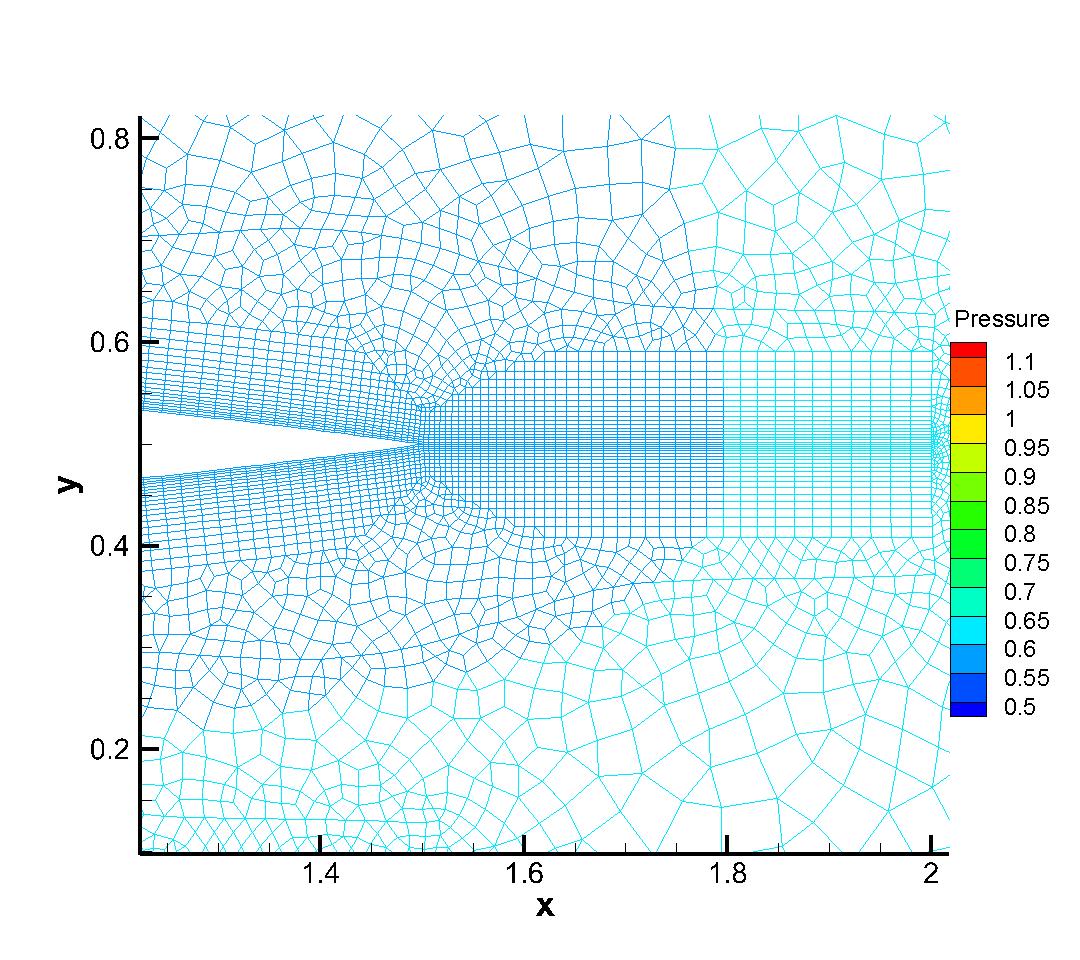}
	\caption{\label{naca0012-dual-mesh}
		Local mesh distributions colored by pressure for the dual NACA0012 airfoil case. Ma=0.8. Re=500. AOA=10$^{\circ}$.}
\end{figure}

\begin{figure}[htp]	
	\centering	
	\includegraphics[height=0.35\textwidth]{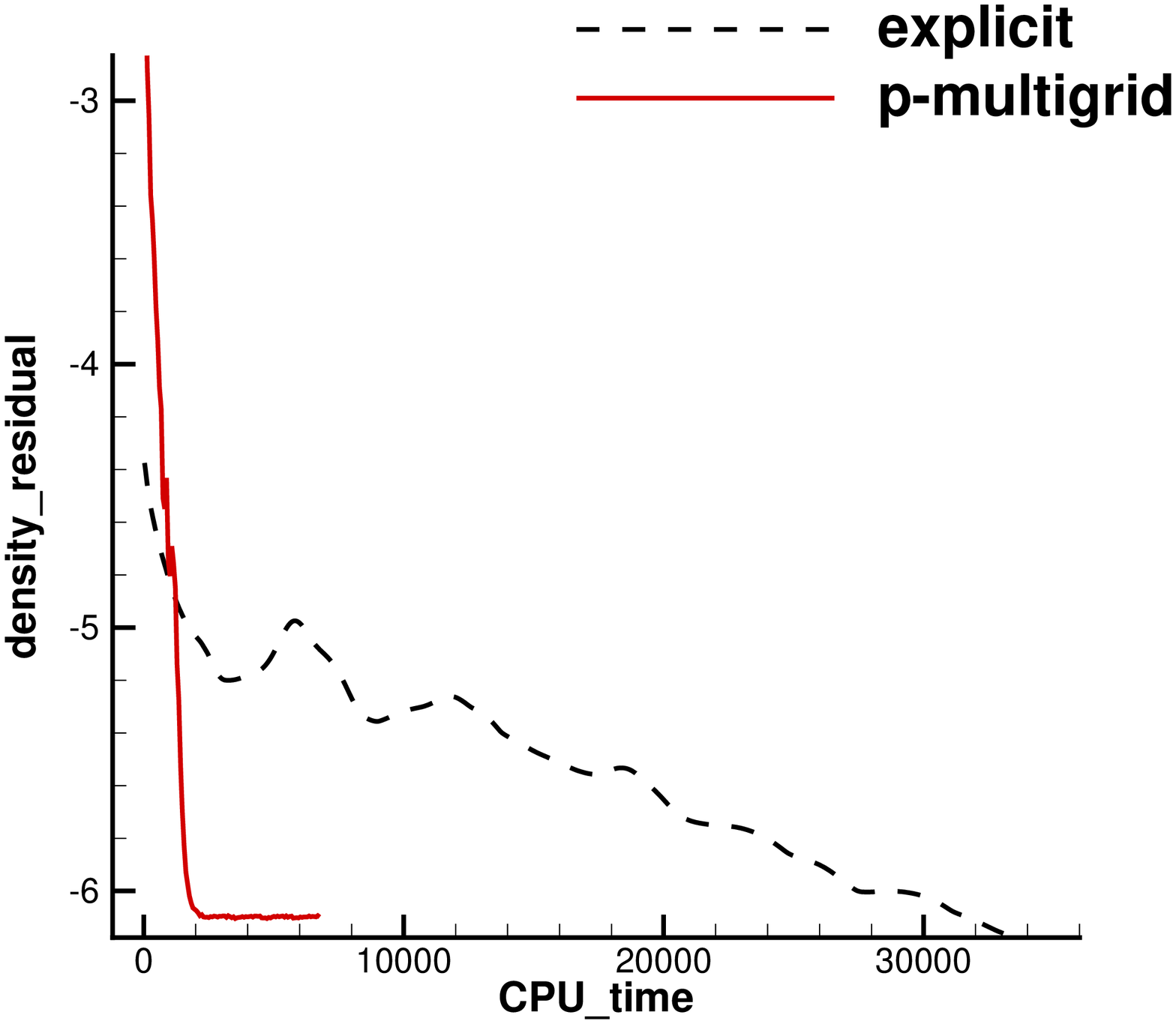}
	\includegraphics[height=0.35\textwidth]{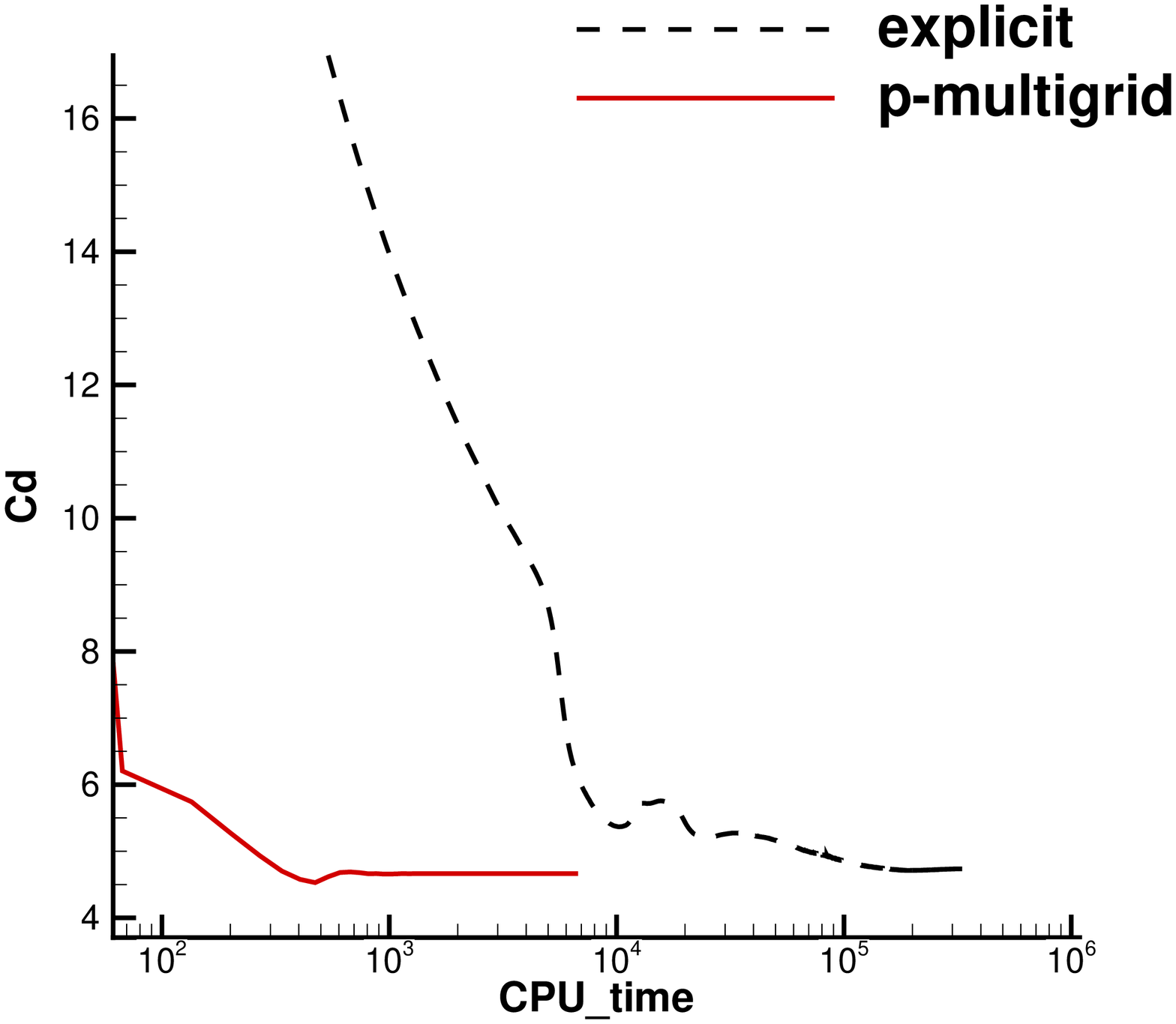}
	\caption{\label{naca0012-dual-res}
		The residuals and total drag coefficients of the dual NACA0012 airfoil cases.}
\end{figure}

\begin{figure}[htp]	
	\centering	
	\includegraphics[width=0.4\textwidth]{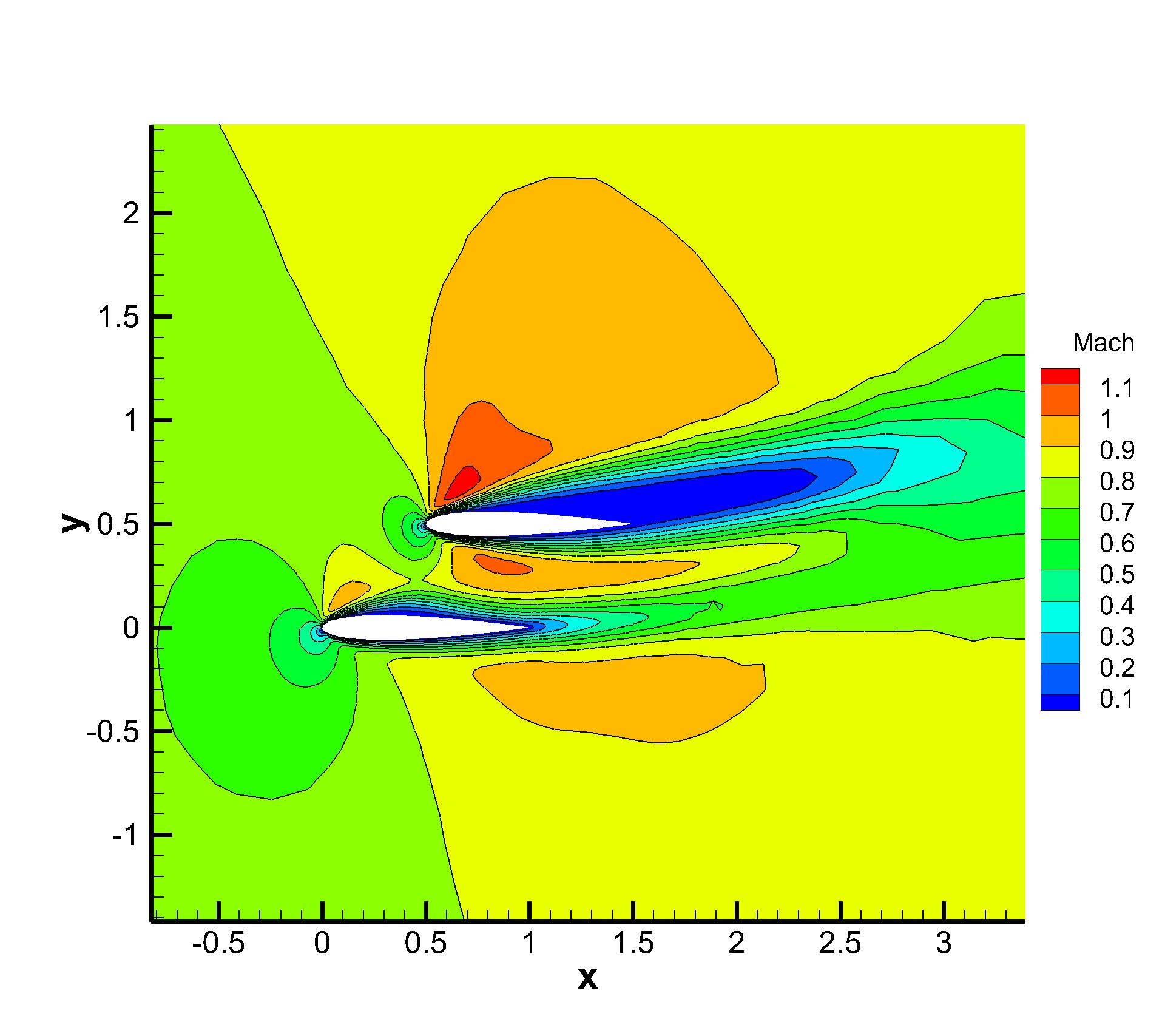}
	\includegraphics[width=0.4\textwidth]{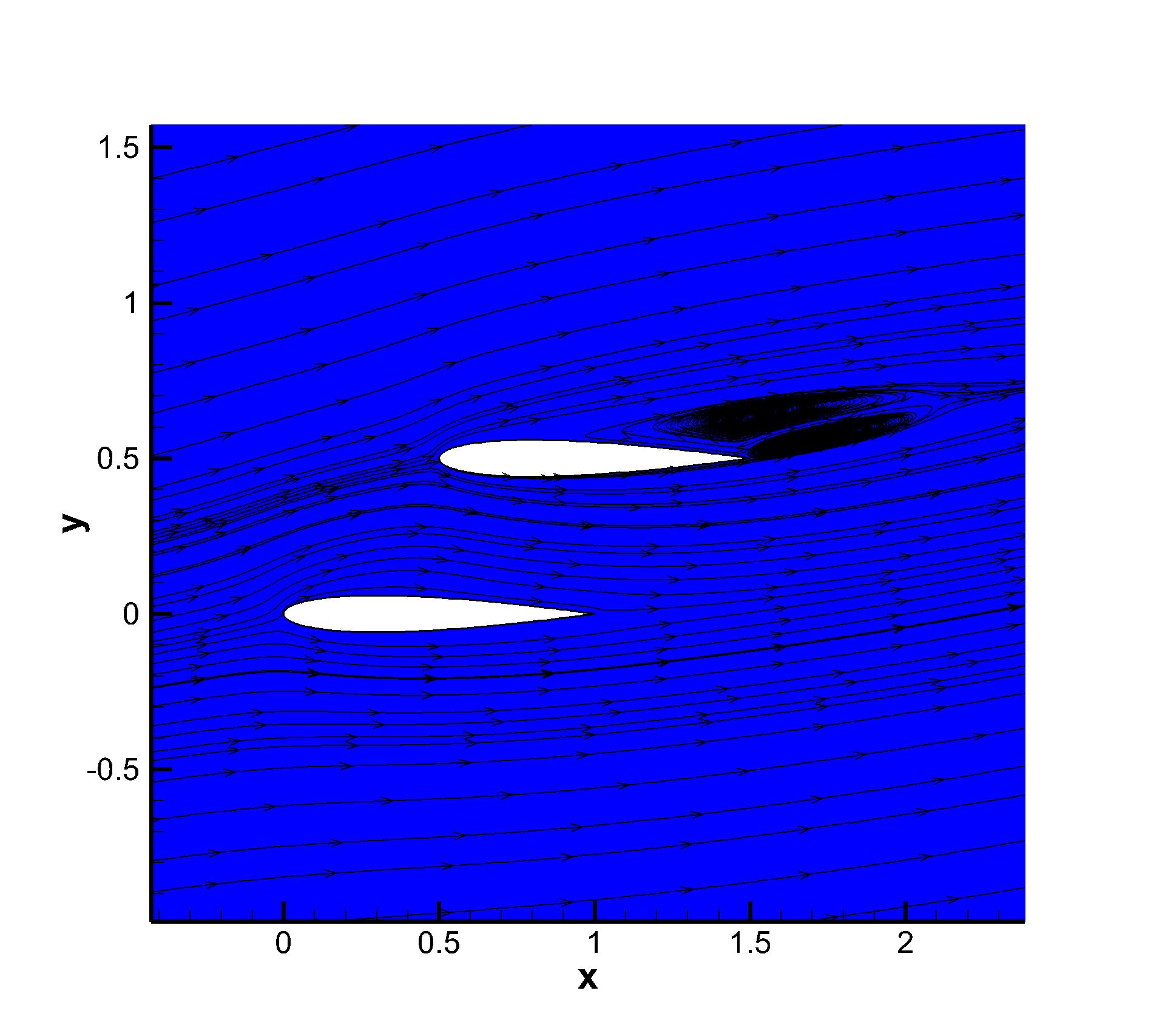}
	\caption{\label{naca0012-dual-contour}
		Transonic flow passing through dual NACA0012 airfoils by the p-multigrid CGKS. Left: Mach distribution. Right: streamline.}
\end{figure}

\begin{figure}[htp]	
	\centering	
	\includegraphics[height=0.35\textwidth]{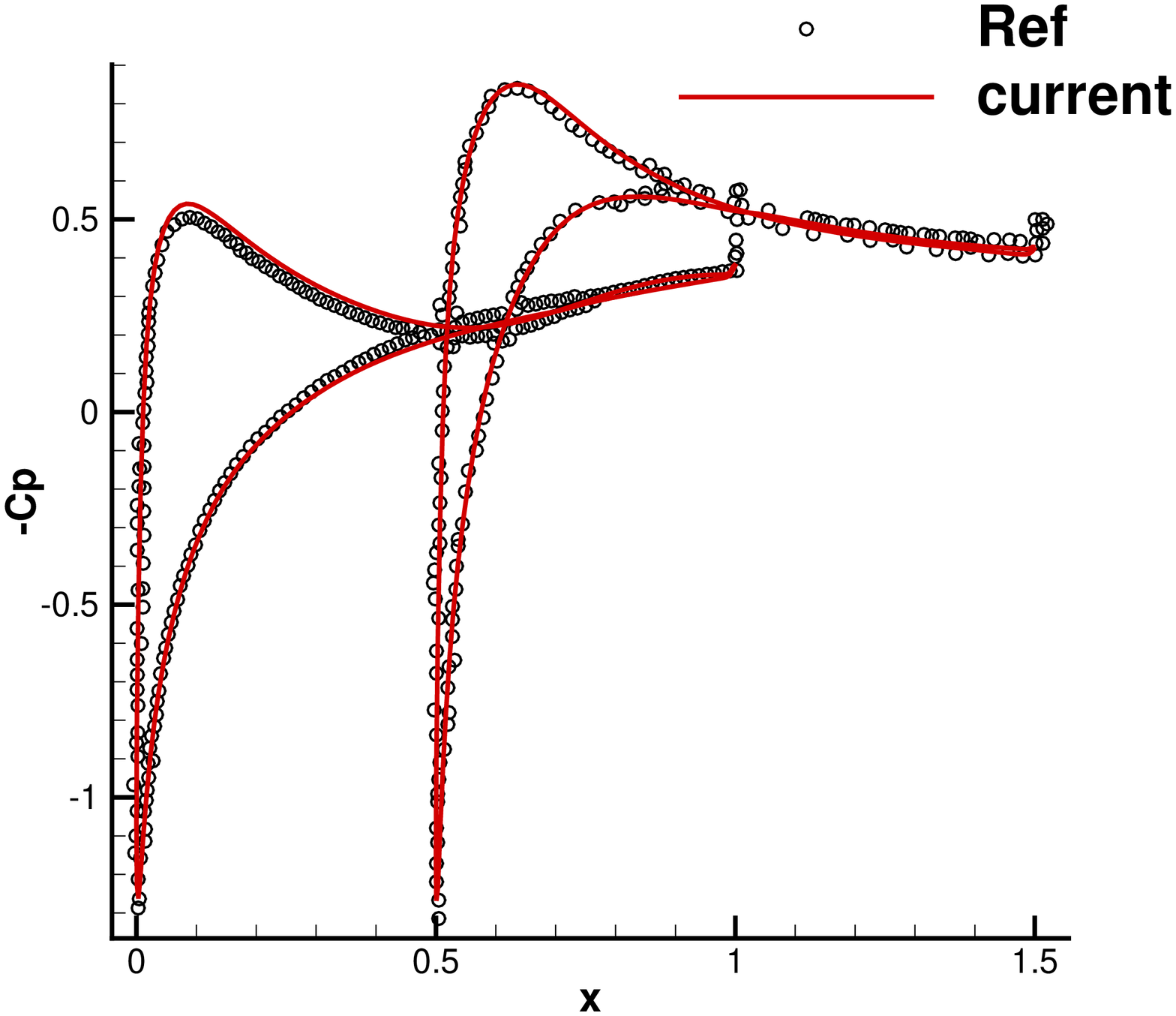}
	\includegraphics[height=0.35\textwidth]{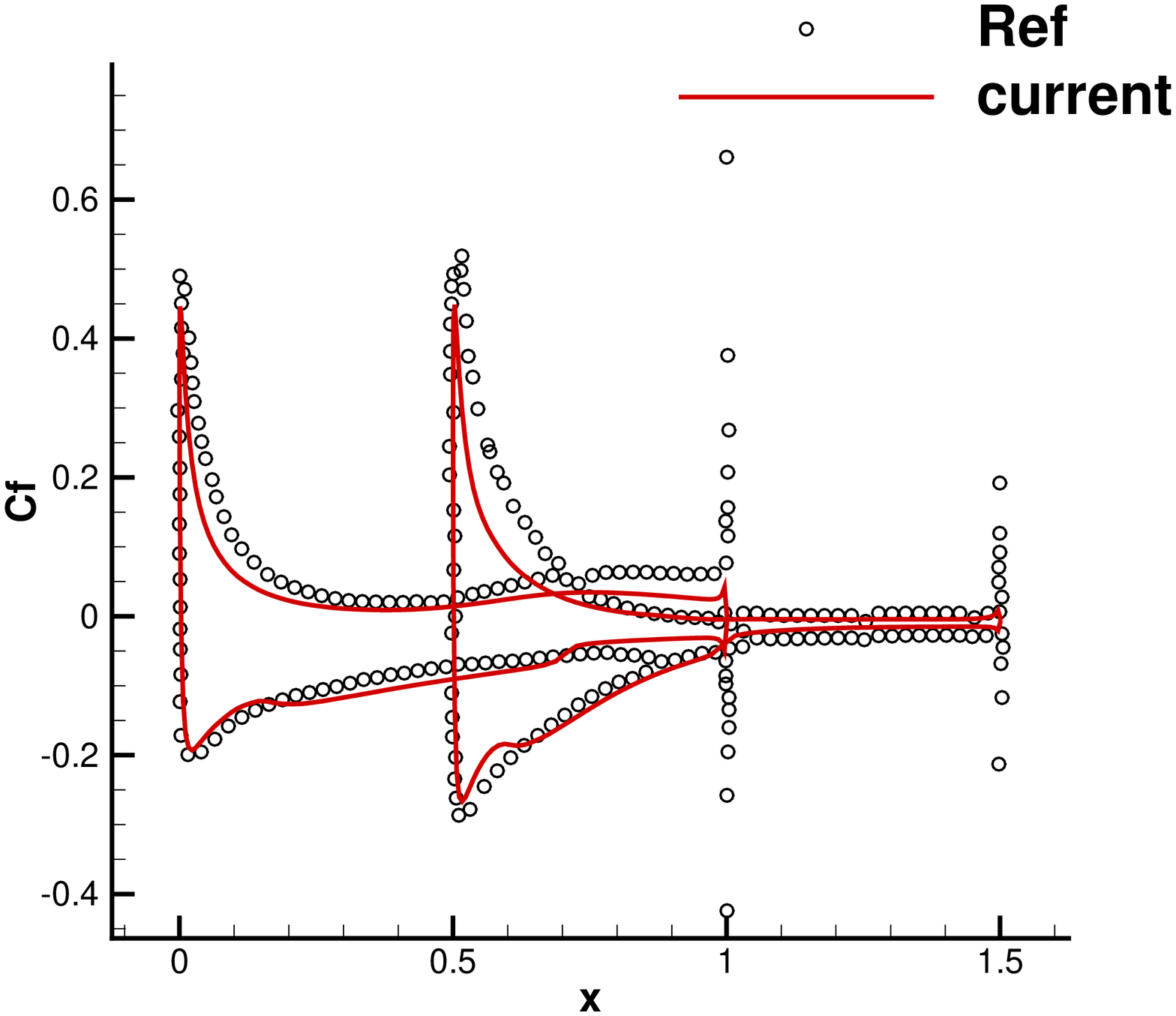}
	\caption{\label{naca0012-dual-line}
		Transonic flow passing through dual NACA0012 airfoils by the p-multigrid CGKS. Left: surface pressure coefficient. Right: skin friction coefficient.}
\end{figure}

To further test the convergence performance for the current CGKS with discontinuities and complex geometry, a two-dimensional transonic flow passing through dual NACA0012 airfoils is tested.
The head of the first airfoil is located at $(0,0)$ and the second one is located at $(0.5,0.5)$.
Both airfoils are put in parallel with the x-axis.
The Reynolds number is Re=500 based on the chord length $L=1$. The incoming Mach number is Ma=0.8 and the AOA is 10$^{\circ}$.
The airfoil is set as non-slip and adiabatic.
Unstructured mixed-element mesh with $14339 \times 2$ cells is used with a near wall size $h=2\times10^{-3}$.
The local enlargements of the mesh and the pressure distributions obtained by the p-multigrid CGKS are presented in Fig.~\ref{naca0012-dual-mesh}.
For this case, the residuals of explicit CGKS cannot reach a steady level even with $10^6$ steps, as shown in Fig.~\ref{naca0012-dual-res}.
Thus, the total drag coefficients Cd are used as the convergence criterion.
While the p-multigrid CGKS can obtain a converged Cd within 2000 seconds, the explicit counterpart needs more than $2\times 10^5$ seconds to achieve a similar result, which means the speed up is more than 100.
The Mach distribution and streamline around both airfoils are shown in Fig.~\ref{naca0012-dual-contour}.
The oblique shock wave can be observed at the front of the top airfoil.
Two vortices are formed in the separated region of the top airfoil,
which agrees well with the referenced results by the second-order finite-volume method \cite{jawahar2000high}.
The surface pressure coefficient and skin friction coefficient are also extracted and compared with the reference data \cite{jawahar2000high}, as shown in Fig.~\ref{naca0012-dual-line}.

\subsection{Flow passing through a sphere}
\begin{figure}[htp]	
	\centering	
	\includegraphics[height=0.35\textwidth]{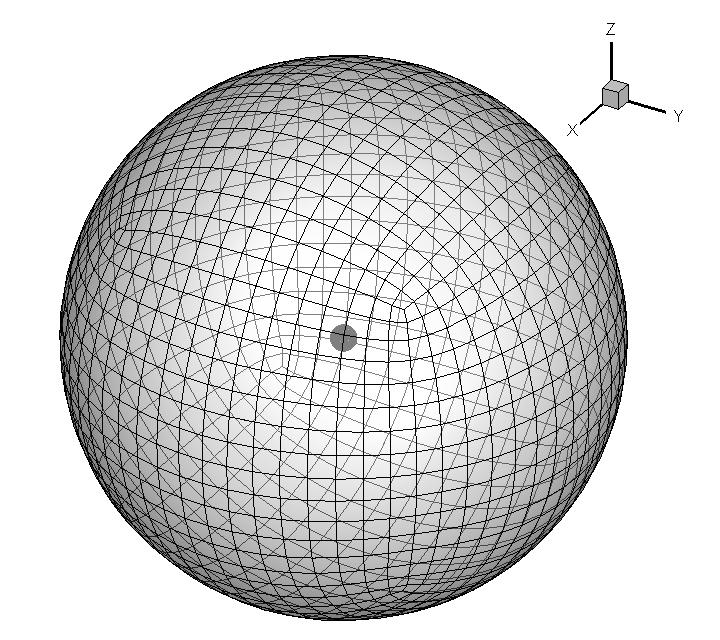}
	\includegraphics[height=0.35\textwidth]{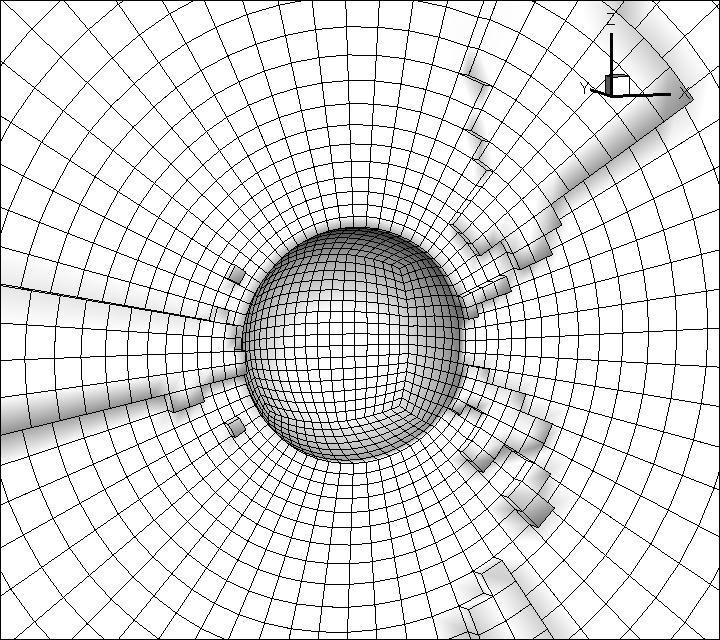}
	\caption{\label{viscous-sphere-mesh}
		Flow passing through a sphere. Mesh sample.}
\end{figure}

Flow passing through a sphere with different Mach number are tested.
The Reynolds number is based on the diameter of the sphere $D=1$.
The far-field condition is set at outside boundary of the domain with the free stream condition
\begin{equation*}
\begin{split}
(\rho,U,V,W,p)_{\infty} =(1,Ma,0,0,\frac{1}{\gamma}),
\end{split}
\end{equation*}
with $\gamma=1.4$.
The non-slip adiabatic boundary condition is imposed on the sphere.
The mesh sample is shown in Fig.~\ref{viscous-sphere-mesh}.

\noindent{\sl{(a) Subsonic case}}

\begin{figure}[htp]	
	\centering	
	\includegraphics[height=0.35\textwidth]
	{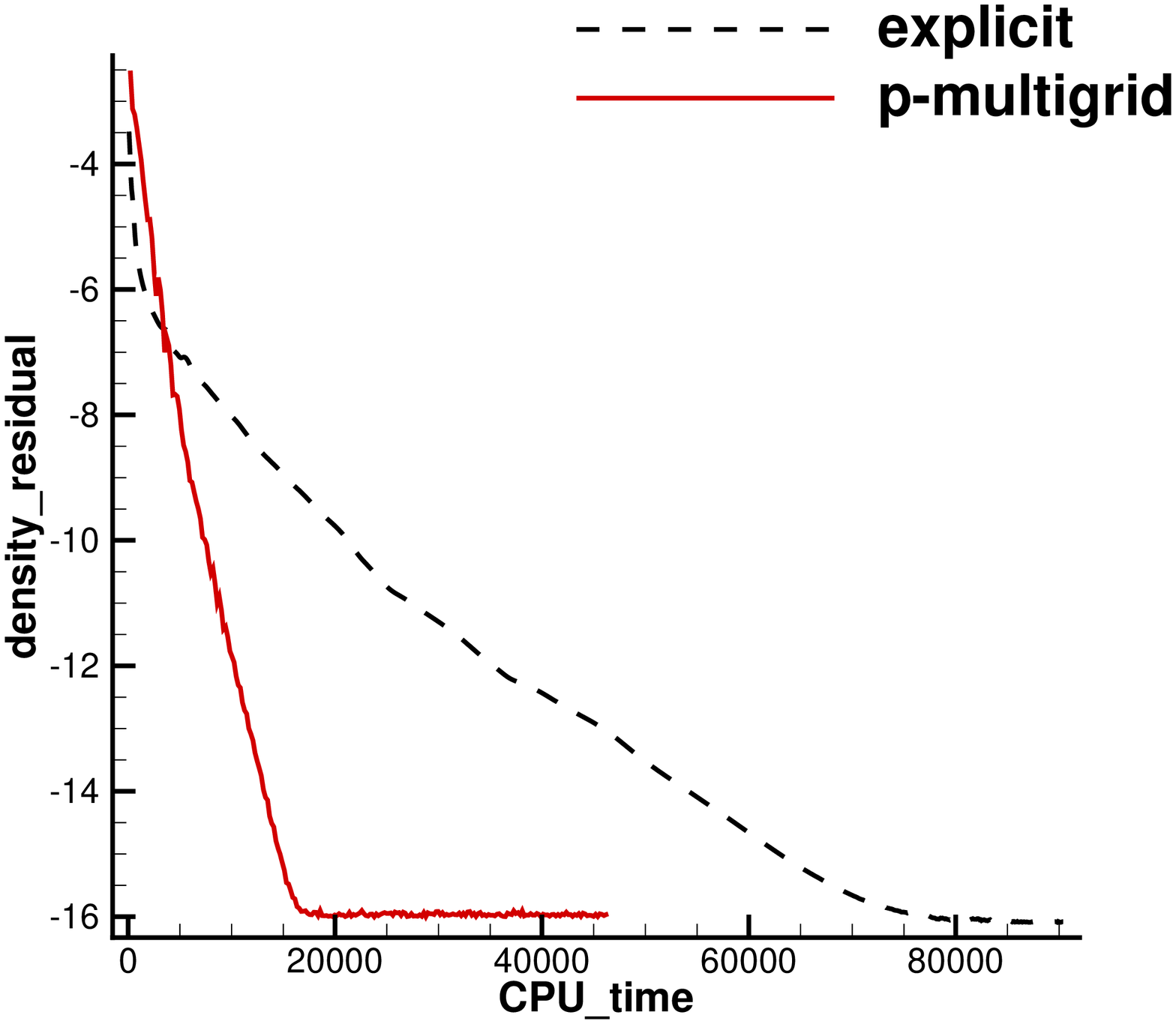}
	\includegraphics[height=0.35\textwidth]
	{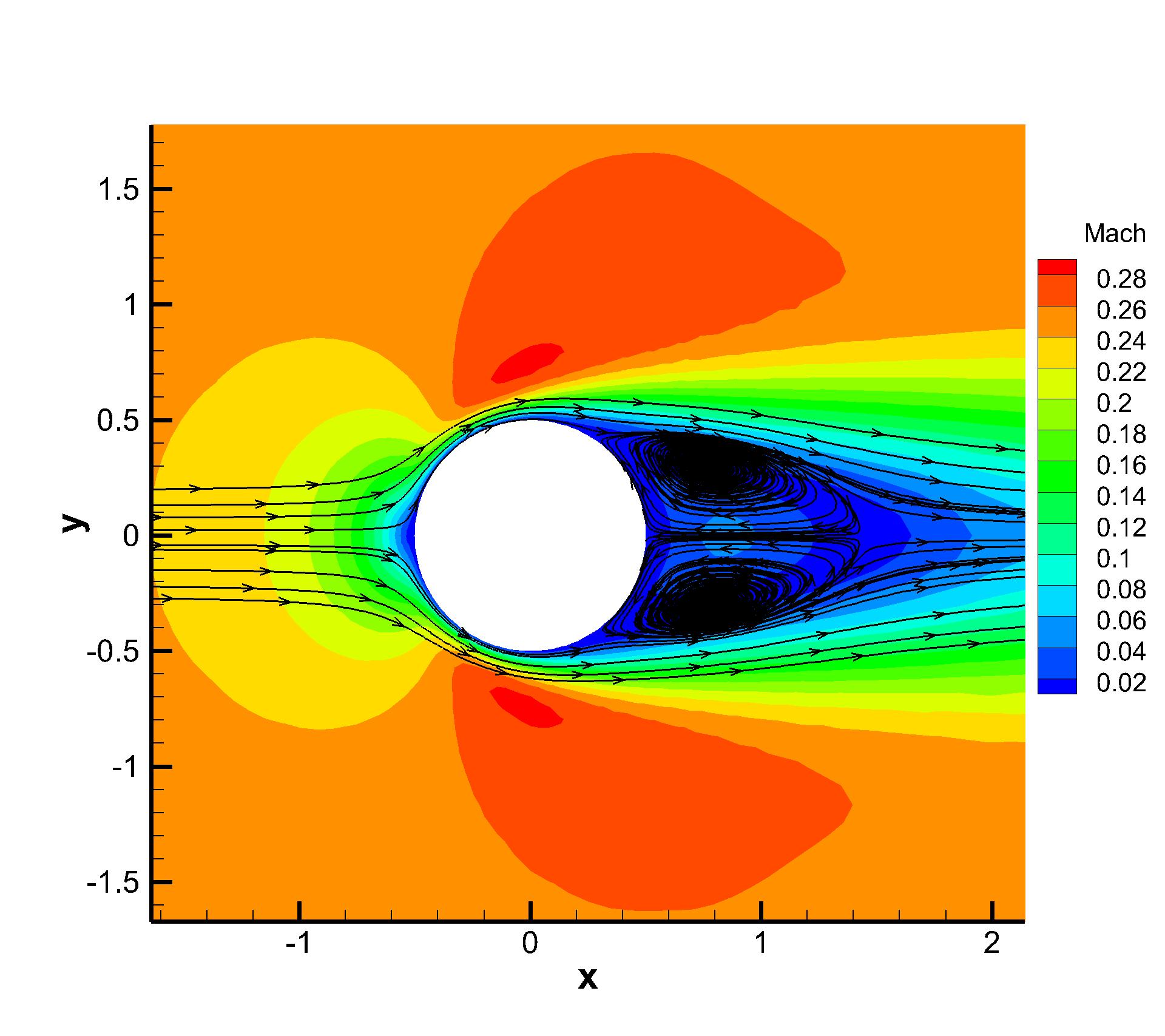}
	\caption{\label{viscous-subsonic-sphere-res}
		Subsonic sphere case. Ma=0.2535. Re=118. Left: density residuals. Right: Mach contour and streamlines obtained by the p-multigrid CGKS. }
\end{figure}

A low-speed viscous case with Re=118 is presented first.
The first mesh off the wall has the size $ h \approx 1.0 \times 10^{-2} D$, and the total CELL number is $6144\times30$.
The residuals can be close to machine zero for both explicit and p-multigrid CGKS.
The speedup for the p-multigrid CGKS is about 4.5 at a residual level of $10^{-16}$, as shown in Fig.~\ref{viscous-subsonic-sphere-res}.
The Mach contour and streamline are also presented in Fig.~\ref{viscous-subsonic-sphere-res} to show the high resolution of the CGKS.
Quantitative results are given in Table \ref{viscous-subsonic-sphere}, including the drag coefficient Cd,  the separation angle $\theta$, and the closed wake length $L$, as defined in \cite{ji2021compact}.

\begin{table}[htp]
	\small
	\begin{center}
		\def\temptablewidth{1.0\textwidth}
		{\rule{\temptablewidth}{1pt}}
		\begin{tabular*}{\temptablewidth}{@{\extracolsep{\fill}}c|c|c|c|c|c}
			Scheme & Mesh number & Cd  & $\theta$  &L &Cl\\
			\hline
			Experiment \cite{taneda1956experimental}	&-- & 1.0  & 151 & 1.07 & -- \\ 	
			Third-order DDG \cite{cheng2017parallel} & 160,868 & 1.016 & 123.7 & 0.96 & --\\
			Fourth-order VFV \cite{wang2017thesis}  & 458,915 & 1.014 & --& -- & 2.0e-5\\
			Current & 184,320 & 1.002  & 124.9 & 0.94 & 3.5e-3\\
		\end{tabular*}
		{\rule{\temptablewidth}{0.1pt}}
	\end{center}
	\vspace{-4mm} \caption{\label{viscous-subsonic-sphere} Quantitative comparisons among different compact schemes  for the subsonic flow passing through a sphere.}
\end{table}

\noindent{\sl{(d) Supersonic case}}

\begin{figure}[htp]	
	\centering
	\includegraphics[width=0.32\textwidth]
    {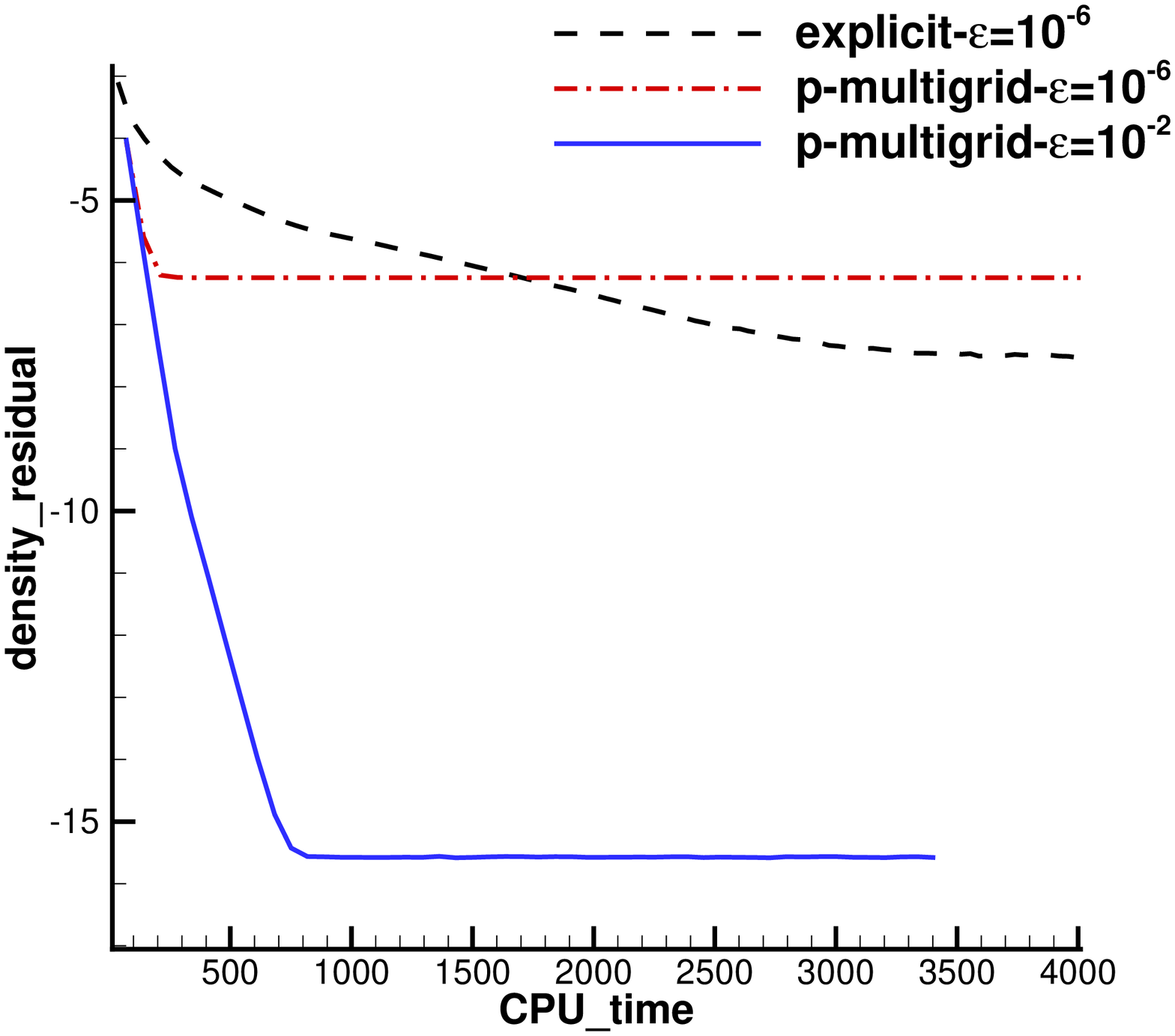}
    	\includegraphics[width=0.32\textwidth]
    {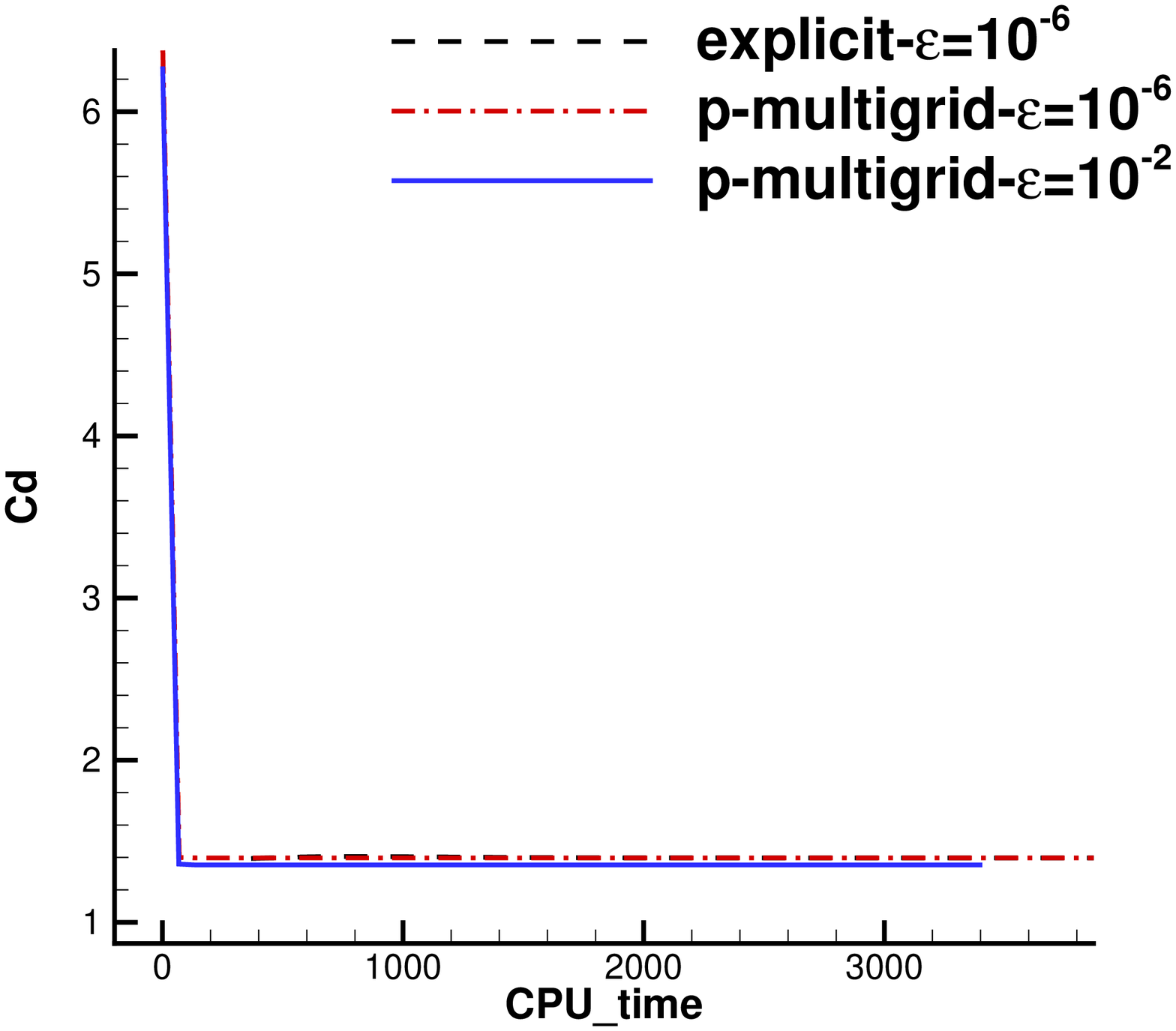}
     	\includegraphics[width=0.32\textwidth]
    {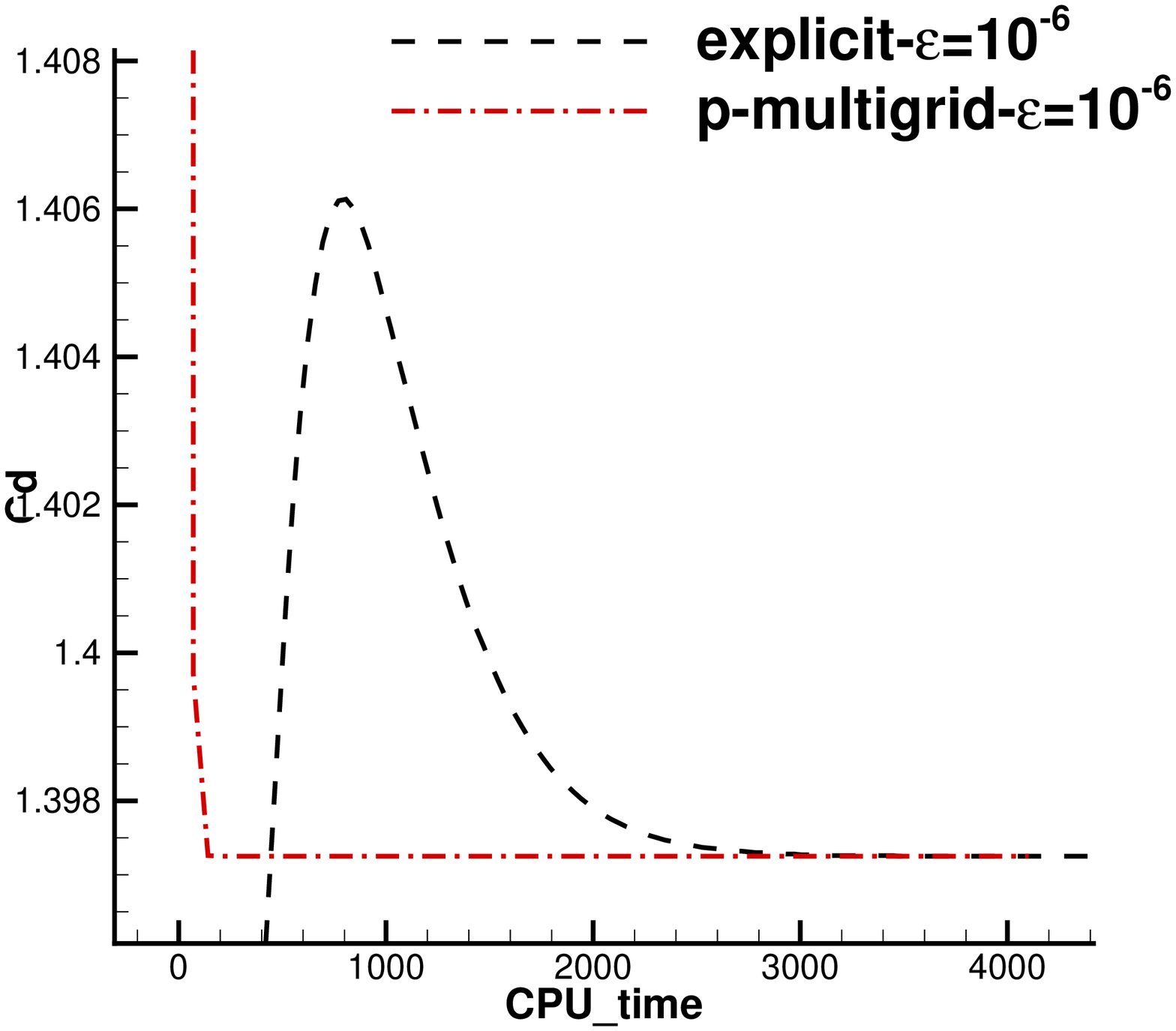}
	\caption{\label{viscous-supersonic-sphere-res}
		The CPU time history of the density residuals and drag coefficients of the supersonic sphere case.}
\end{figure}

\begin{figure}[htp]	
	\centering
	\includegraphics[width=0.32\textwidth]
	{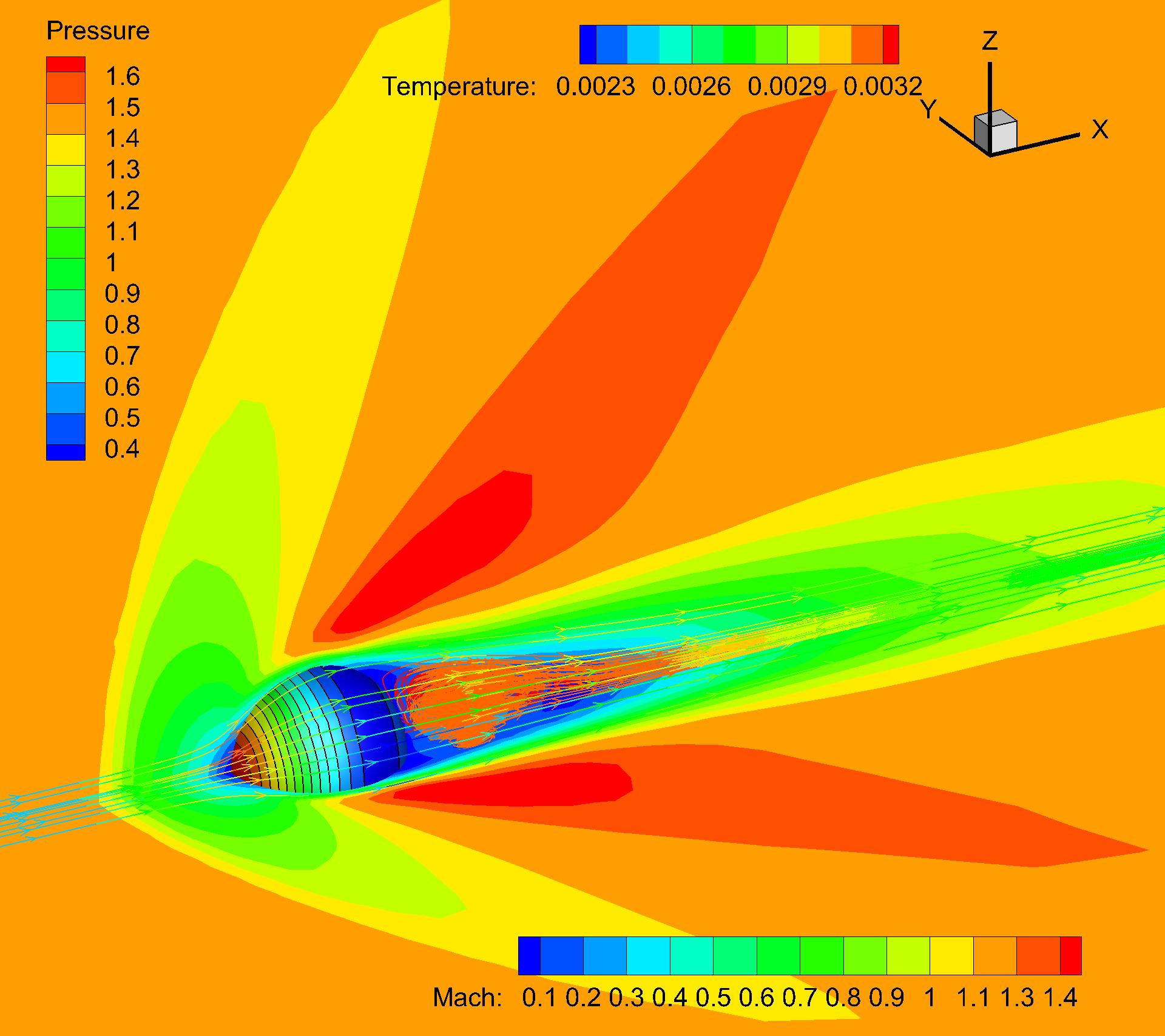}
	\caption{\label{viscous-supersonic-sphere-ma}
		Supersonic flow passing through a viscous sphere by the p-multigrid CGKS. Ma=1.2. Re=300.}
\end{figure}

To validate the effectiveness of the current scheme for the 3-D high-speed viscous flow, a supersonic viscous sphere with Ma=1.2 is tested.
The Reynolds number is set as 300 and the Prandtl number is $Pr=1$.
The first mesh off the wall has the size $ h \approx 4.5 \times 10^{-2} D$
and the total cell number is 50688.
Fig.~\ref{viscous-supersonic-sphere-res} shows the density residuals of different schemes.
Similar to the shock reflection problem, the small parameter $\epsilon$ can significantly affect the final convergent residuals.
However, similar drag coefficients can be obtained for different $\epsilon$, with a relative error of about 3\%.
With the same $\epsilon$, the almost identical drag coefficients are obtained for both the p-multigrid and explicit CGKS.
The p-multigrid CGKS obtains a converged Cd within 200 seconds, where a more than 15 speedup is achieved in comparison with the explicit one.
The numerical results including the Mach contour, surface pressure distribution, and streamline around sphere are shown in Fig.~\ref{viscous-supersonic-sphere-ma}.
Quantitative results are listed in Table \ref{viscous-supersonic-sphere},
which have a good agreement with those given by Nagata et al. \cite{Nagata2016sphere}.

\begin{table}[htp]
	\small
	\begin{center}
		\def\temptablewidth{1.0\textwidth}
		{\rule{\temptablewidth}{1pt}}
		\begin{tabular*}{\temptablewidth}{@{\extracolsep{\fill}}c|c|c|c|c|c}
			Scheme & Mesh Number & Cd  & $\theta$  &L & Shock stand-off\\
			\hline
			WENO6 \cite{Nagata2016sphere} 	&909,072 & 1.281  & 150.9 & 0.38 & 0.21 \\ 	
			Current with $\epsilon=10^{-6}$ & 50,688 & 1.398  & 148.5 & 0.45 & 0.28-0.31 \\
			Current with with $\epsilon=10^{-2}$ & 50,688 & 1.354  & 149.2 & 0.45 & 0.28-0.31 \\	
		\end{tabular*}
		{\rule{\temptablewidth}{0.1pt}}
	\end{center}
	\vspace{-4mm} \caption{\label{viscous-supersonic-sphere} Quantitative comparisons between the current scheme and the reference solution for the supersonic flow passing through a sphere.}
\end{table}

\section{Conclusions}
In this paper, the p-multigrid method is adopted for the CGKS.
A two-level algorithm is used to drive the scheme to the convergent steady-state solutions.
In the high-order level, the explicit single-step third-order CGKS is used.
In the low-order level, the first-order scheme with implicit point-relaxation method is  applied as an iterative smoother.
The resulting scheme is simple and low-storage.
A series of test cases is presented on 3-D mixed-element mesh.
It is worth remarking that the residuals change almost identically along with the iteration steps under either serial or parallel computation, which suggests the current scheme is highly scalable.
Efficiency in aspect of CPU time can be increased by one order of magnitude for the p-multigrid CGKS in comparison with the original explicit scheme.
The numerical results suggest that higher speedup can be achieved under the mesh with a wide variation of cell size.
The algorithm is not sensitive to the use of difference flux solvers on the low-order level.
Moreover, the convergence for transonic and supersonic flow is studied.
The quantitative results from the p-multigrid method are almost identical to the explicit counterpart.
The same order of speedup can be kept for high-speed flow.
In further work, the CGKS coupled with turbulent modeling will be developed for the Reynolds-averaged Navier-Stokes solutions.

\section*{Acknowledgments}

The authors would like to thank Dr. Yajun Zhu and Mr. Yue Zhang for helpful discussions on the p-multigrid method.
The current research is supported by National Numerical Windtunnel project,  National Science Foundation of China (11772281, 91852114),
Hong Kong Research Grant Council (16208021). 

\section*{References}
\bibliographystyle{plain}%
\bibliography{jixingbib}

\end{document}